\documentclass[aps,pre,twocolumn,superscriptaddress,showpacs]{revtex4}
\usepackage{graphicx}
\usepackage{amssymb}
\begin{document}
\bibliographystyle{apsrev}

\title{Contact and Friction of Nano-Asperities: Effects of Adsorbed Monolayers}
\author{Shengfeng Cheng}
\affiliation{Department of Physics and Astronomy, Johns Hopkins University,\\
3400 N Charles St, Baltimore, MD 21218, USA}
\author{Binquan Luan}
\affiliation{IBM T. J. Watson Research Center, P.O. Box 218, Yorktown Heights, NY 10598, USA}
\author{Mark O. Robbins}
\affiliation{Department of Physics and Astronomy, Johns Hopkins University,\\
3400 N Charles St, Baltimore, MD 21218, USA}

\date{\today}

\begin{abstract}
Molecular dynamics simulations are used to study contact between
a rigid, nonadhesive, spherical tip with radius of order $30$nm and
a flat elastic substrate covered with a fluid monolayer of adsorbed
chain molecules.
Previous studies of bare surfaces showed that the atomic scale deviations
from a sphere that are present on any tip constructed
from discrete atoms lead to significant deviations from continuum theory
and dramatic variability in friction forces.
Introducing an adsorbed monolayer leads to larger deviations from
continuum theory,
but decreases the variations between tips with different atomic structure.
Although the film is fluid, it remains in the contact and behaves
qualitatively like a thin elastic coating except for
certain tips at high loads.
Measures of the contact area
based on the moments or outer limits of the pressure distribution
and on counting contacting atoms are compared.
The number of tip atoms making contact in a time interval grows as a power
of the interval when the film is present and logarithmically with the
interval for bare surfaces.
Friction is measured by displacing the tip at a constant velocity 
or pulling the tip with a spring. 
Both static and kinetic friction rise linearly with load at small loads.
Transitions in the state of the film lead to nonlinear behavior at large loads.
The friction is less clearly correlated with contact area than load.
\end{abstract}

\pacs{46.55.+d,~62.20.Qp,~81.40.Pq}

\maketitle

\section{Introduction}

Studies of the contact between a spherical tip and a flat substrate have
played a
central role in models of the mechanical and frictional response of surfaces.
In continuum mechanics, this geometry can be mapped into contact between
two peaks or asperities on opposing rough surfaces as they are pressed
together~\cite{johnson85}.
The response of the entire system is then commonly
represented as a sum over many
independent contacts~\cite{greenwood66,johnson85,bowden86},
although recent studies emphasize the importance of interactions
between asperities~\cite{persson01,hyun04,campana08,persson08c}.

Ideal single asperity contacts are difficult to realize in experiments because
both the tip and substrate are usually rough
on scales smaller than the average radius of curvature.
As a result, most experiments with micrometer and larger tips end up measuring
the response of many smaller asperities.
Notable exceptions are experiments on atomically flat
mica surfaces in the Surface Force Apparatus~\cite{gee90,granick91,klein95}.
The main limitation of these experiments is that the contact pressure
is typically only tens of megapascals, which is orders of magnitude
smaller than estimates of the pressure between asperities on rough 
surfaces.

In the last twenty years there has been great interest in the study
of single asperities at the nanometer scale using the atomic force
microscope (AFM) and related scanning probes~\cite{carpick97,kiely98,bhushan04}.
When the AFM is operated in a typical contact mode, a tip with
characteristic radius of $10$ to $100$nm is pressed into contact
with a surface and may be translated to measure the friction.
The small dimensions of the tip allow more direct control and/or
measurement of the chemistry and geometry, and reduce the range
of roughness that can exist on the surfaces.

AFM experiments provide direct information about the normal and lateral
forces, but do not reveal the distribution of forces within the contact,
the area of contact, or the mechanisms of deformation and friction.
This prevents them from directly answering long standing questions about
the role of area and load in determining friction.
At the macroscopic scale, Amontons's laws of friction say that the
friction is proportional to load and independent of the area of contact.
However a prevailing view since the mid 1900's has been that the
friction is proportional to the real area of molecular contact $A_{real}$, 
which is much smaller than the nominal 
or apparent area~\cite{bowden86,johnson85,greenwood66}.
The real area typically grows linearly with load for nonadhesive
surfaces, and Amontons's laws are recovered if there is a constant
shear stress $\tau_{shear}$ in the contact.
For adhesive surfaces, $A_{real}$ remains finite at zero load, explaining
why friction is often finite at zero load.

To try to address the role of $A_{real}$ at the nanometer scale,
AFM experiments have been compared to continuum models.
The measured friction can be fit by assuming both that continuum
theory describes $A_{real}$ in nanometer scale contacts
and that there is a constant $\tau_{shear}$~\cite{carpick97,carpick99,schwarz03}.
However, the success of such fits need not imply that the underlying assumptions
are valid, because the functional dependence is quite simple and there are
many poorly constrained fit parameters.
This has been demonstrated in subsequent simulations where the friction
can be fit in the same way, 
but both assumptions break down~\cite{luan05,luan06b,szlufarska09}.
The major cause of the discrepancy with continuum theory is the
atomic scale roughness that is present on any surface composed of
discrete atoms.
This spreads the contact over a larger area, much as predicted by
continuum calculations for rough surfaces~\cite{johnson85,greenwood67}.

These simulations of single asperity contacts between clean
surfaces have found a variety of relations between $A_{real}$,
load and friction.
The friction changes by more than two orders of magnitude with the
precise atomic structure of the tip~\cite{luan05}, and
may scale linearly with area, load or
neither~\cite{luan05,luan06b,szlufarska09}.
Similar strong dependence on atomic structure is found for the
friction between flat surfaces, with friction expected to
vanish with increasing area unless the surfaces
are commensurate, i.e., share a common 
period~\cite{hirano90,muser01tl,muser01prl,hirano06,dienwiebel04,dietzel08,martin93}.

Of course experimental surfaces are rarely clean.
Any surface exposed to air is rapidly coated with adsorbed molecules
of water, oxygen or small hydrocarbons.
Wear debris, dust and larger particles may also be present.
These ``third bodies'' can have a profound effect on friction.
Studies of flat surfaces show that they greatly reduce the sensitivity
of friction to atomic structure and produce a friction force that
rises linearly with normal load~\cite{he99,he01tl,he01b,muser01prl}.

In this paper we consider the effect of adsorbed monolayers on the
mechanical properties and friction of single asperity contacts.
Even though the adsorbed molecules freely diffuse along the surface,
they are trapped in a glassy state when confined under the tip
and remain in the contact at very high loads.
As for flat surfaces, the adsorbed monolayer decreases the variation
of friction with atomic structure and yields a force that usually
rises linearly with normal load.
The adsorbed monolayer also decreases variations with
tip structure in the contact area and normal stiffness,
but leads to large
deviations from the results for bare tips.
The observed increase in contact area and decrease in stiffness
relative to continuum theory are qualitatively
consistent with modeling the adsorbed monolayer as a
thin elastic coating.

The paper is organized as follows.
Sec.~II describes the interactions, geometry and methods used
in our simulations.
The MD results for the pressure distribution, contact radius,
tip displacement, and friction forces are presented in Sec.~III.
Sec.~IV provides a brief summary and discussion of the results.

\section{Simulation Methods}
\begin{figure}[htb]
\centering
\includegraphics[width=3in]{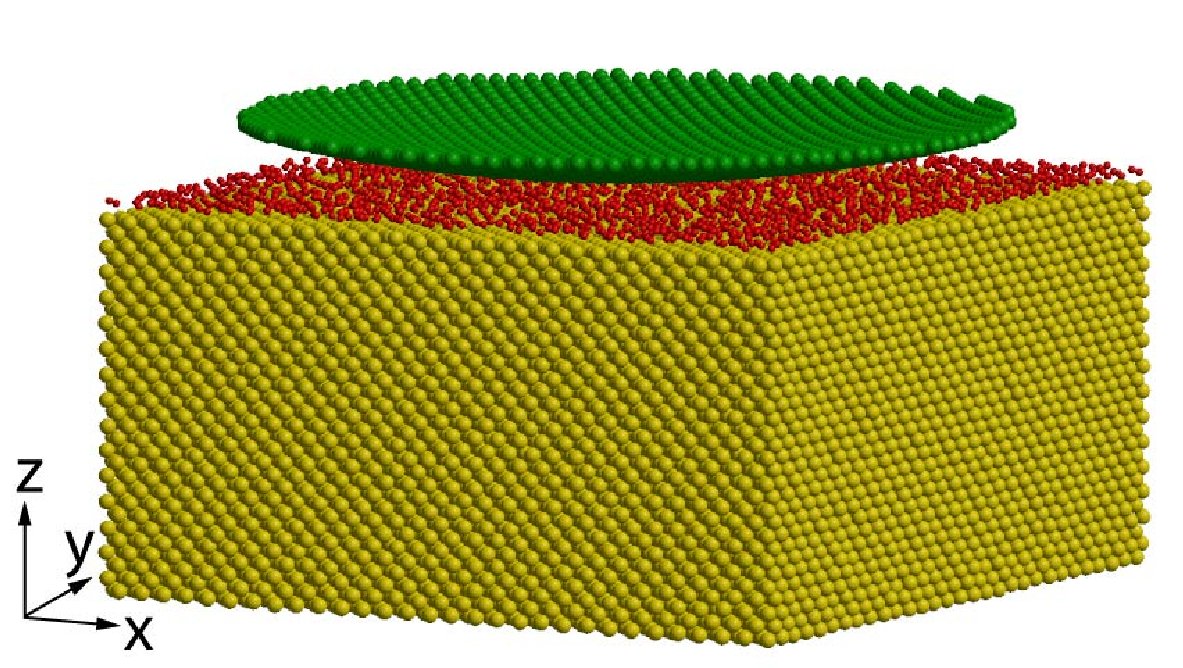}
\caption{(color online) A tip (top) in contact with a substrate (bottom) 
covered with an adsorbed fluid layer (red).}
\label{AtomicConfig}
\end{figure}

Figure \ref{AtomicConfig} shows a snapshot of the system studied in our MD simulations.
A spherical tip is pushed, under a controlled normal load, 
into a flat substrate covered with an adsorbed fluid layer.
The picture only shows a small region near the center of contact
(about $\frac{1}{8}\times\frac{1}{8}$ in the plane of the substrate surface 
and $\frac{1}{10}$ in the substrate depth).
The tip radius is $R \sim 30 {\rm nm}$, which is comparable to radii of tips used in AFM experiments.
The geometry also approximates contact of peaks on randomly rough surfaces.

In continuum mechanics a contact between two elastic, frictionless spheres,
with radii $R_1$ and $R_2$, Young's moduli $E_1$ and $E_2$, 
and Poisson ratios $\nu_1$ and $\nu_2$ respectively,
can be mapped to a contact between a rigid sphere of radius
$R=(R_1^{-1}+R_2^{-1})^{-1}$ and a flat elastic solid 
of modulus $E^{*}$ given by~\cite{johnson85}
\begin{equation}
\label{EffectiveModulus}
1/E^{*}=(1-\nu_1^{2})/E_1+(1-\nu_2^{2})/E_2~.
\end{equation}
Previous studies have shown that the mapping remains essentially 
correct at atomic scales~\cite{luan05}. 
Thus simulations can be simplified to the case of a rigid tip and 
an elastic substrate without loss of generality.

To approximate a perfectly elastic material, we take a face-centered-cubic (fcc) crystal 
as the substrate with a (001) surface exposed. 
Atoms on nearest-neighbor sites in the substrate are connected by a harmonic potential
\begin{equation}
\label{SpringPotential}
V_{\rm S}(r) = \frac{1}{2}k(r-d)^{2},
\end{equation}
where $k$ is a spring constant, $d$ is the equilibrium distance between two nearest neighbors,
and $r$ is the inter-atomic distance. To be consistent with our previous work,
we express $k$ and $d$ in terms of the values for particles interacting with a Lennard-Jones (LJ) potential.
The truncated and shifted LJ potential is
\begin{equation}
 \label{LJPotential}
V_{LJ}(r)=
4\epsilon\left[ \left(\frac{\sigma}{r}\right)^{12}-\left(\frac{\sigma}{r}\right)^{6}-
\left(\frac{\sigma}{r_c}\right)^{12}+\left(\frac{\sigma}{r_c}\right)^{6} \right]
\end{equation}
for $r \leq r_c$, 
where $\epsilon$ and $\sigma$ are the characteristic binding energy and diameter,
and $r_c$ is the cutoff distance.
If atoms in the solid interacted with this potential and the cutoff only included nearest-neighbors,
then one would have $d = 2^{1/6}\sigma$ and $k$ approximately $57\epsilon/\sigma^2$.
We use these values for the spring interaction described in Eq.~(\ref{SpringPotential})
and express all quantities in terms of $\sigma$, $\epsilon$ and the mass of solid atoms $m$.
For example, the unit of time is $\sqrt{m\sigma^2/\epsilon}$.
The relation between the effective modulus in Eq.~(\ref{EffectiveModulus})
and the elastic constants is complicated when the solid is cubic rather than isotropic~\cite{johnson85}.
The value $E^{*}=63\epsilon/\sigma^3$ was obtained from fits to
simulation results for the normal displacement in contacts between
a bare fcc crystal and a dense tip
where continuum theory is most accurate~\cite{luan06b}.

Periodic boundary conditions are imposed in the $x$ and $y$ directions, 
i.e., in the substrate surface plane.
The bottom layer of substrate atoms is held fixed.
This mimics the external support that balances the normal load applied to the tip.
Continuum solutions assume a semi-infinite substrate.
To approximate this, all dimensions of the substrate should be much bigger than the contact radius.
The size of the substrate used in this paper is about $195\sigma \times 195\sigma \times 190\sigma$,
which corresponds to a total of more than $6.9$ million atoms.
The contact radius in our MD simulations, 
as measured by the second moment of the pressure distribution,
is less than $7\%$ of the substrate dimensions.
The quantity most affected by finite substrate size is the normal 
displacement.
The magnitude of the corrections to the Hertz prediction has been obtained
by fitting continuum calculations~\cite{johnson01,sridhar04,adams06} 
to simulations for bare contacts~\cite{luan06b}.
The corrections are less than $5\%$ and have been included in
the Hertz prediction lines for normal displacement below.

We perform a series of simulations using tips with different atomic scale structures and roughness, 
but with the same radius ($R=100\sigma \sim 30{\rm nm}$).
The smoothest tips are made by bending the (001) surface of an fcc crystal
with nearest-neighbor spacing $d^\prime$ into a spherical shape.
The friction between bare surfaces depends on whether $d^\prime / d$ is rational or irrational.
These cases are referred to as commensurate and incommensurate, respectively.
The highest friction occurs for commensurate surfaces 
with the same lattice constant and orientation, i.e., $d^\prime / d=1$.
In this commensurate case, we find that the relative position of the tip 
on the substrate surface plays an important role.
Two extreme cases are investigated here.
When tip atoms are directly above interstitial positions on the substrate surface,
we say the tip is in registry since it represents a continuation of the crystalline structure of the substrate.
On the other hand, when tip atoms are directly above atomic equilibrium positions on the substrate surface, 
we say the tip is out of registry.

Contacting experimental surfaces are rarely in perfect registry and alignment.
Any mismatch between lattice constants ($d$ and $d^\prime$) or misalignment
between lattice orientations leads to incommensurate contact.
To model such cases, we make an incommensurate tip by bending a fcc crystal with $d^\prime / d=1.12342$.
To most closely approximate a continuum, we also consider
a dense tip made by bending a (001) face of a fcc crystal with $d^\prime / d=0.05$ or $0.1$.
The two give nearly identical results for contact area and normal displacements.
Frictional forces are small for dense tips and are not reported because such disparate
lattice constants on opposing surfaces are hard to realize with real atoms.

The tips used in AFM experiments are unlikely to have the geometry of bent crystals.
To generate more realistic tip geometries we take a crystalline or amorphous
material and remove all atoms outside a sphere of radius $R$.
The crystalline state is chosen to be commensurate with $d'/d=1$.
The amorphous state is obtained 
by quenching a fluid state with number density $1.0\sigma^{-3}$.
While all tips considered
here deviate from an ideal sphere by less than an atomic diameter,
the effective roughness is quite different.
The cut crystal has terraced steps on the surface and will be
referred to as the ``stepped'' tip in this paper.
The amorphous surface does not have terraces,
but has fluctuations in height and density (see Fig.~1 of Ref.~\cite{luan06b}).

As for the bent commensurate tip, results for the stepped tip depend on registry.
For most of the results presented in this paper the stepped tip is out-of-registry,
with the atoms in the bottom terrace directly above the
equilibrium positions of surface substrate atoms. 
Another case was considered in simulations of friction, because sliding
along the $x$ axis brings commensurate tips between the in and out of
registry configurations.
The tip was shifted diagonally by $0.5 d$ along both $x$ and $y$.
Tip atoms then slide along a line that is centered between substrate atoms.
This did not change static contact properties significantly, but
lowered the friction by a factor of $2$.

Atoms in the tip ($t$) and substrate ($s$) interact via 
the LJ potential described in Eq.~(\ref{LJPotential})
with $\epsilon_{st}=1.0\epsilon$, $\sigma_{st}=1.0\sigma$, and $r_c=2^{1/6}\sigma$. 
The choice of $r_c$ makes the interaction purely repulsive.
Therefore no adhesion is included. 
Possible effects of adhesion will be explored in future studies.
Except as noted, the adsorbed film prevents direct contact between the tip and substrate.

The adsorbed monolayer is composed of chain molecules containing four atoms.
Here ``atom'' is used in a designative sense and each atom may represent a monomer of an oligomer chain.
Atoms ($a$) that are not connected by a covalent bond interact via the LJ potential with
$\epsilon_{aa}=0.25\epsilon$, $\sigma_{aa}=1.0\sigma$, and $r_c=1.8\sigma$.
Reducing the binding energy by a factor of $4$ 
relative to the substrate ensures that the molecules melt at a lower temperature.
For adjacent atoms in an adsorbed molecule, 
the LJ potential is truncated at $r_c=2^{1/6}\sigma$, and the covalent bond is modeled by
the finite extensible nonlinear elastic (FENE) potential~\cite{kremer90}
\begin{equation}
\label{FENEPotential}
V_{B}(r)=\left\{
\begin{array}{cc}
-\frac{1}{2} K R_0^2 {\rm ln}[1-(r/R_0)^2] \ \ & \ \ {\rm if} \ \ r \leq R_0,\\
\infty \ \ & \ \ {\rm if} \ \ r > R_0.
\end{array}
\right.
\end{equation}
Here $K=30\epsilon_{aa}/\sigma^2=7.5\epsilon/\sigma^2$ is a spring constant 
and $R_0=1.5\sigma$ is a length scale.
Previous studies have shown that this coarse-grained model
provides a good description of short-chain hydrocarbon molecules~\cite{kremer90}, 
and it has been used successfully to model 
a wide range of equilibrium and nonequilibrium behavior~\cite{rottler02,rottler02b,hoy07}.

The monolayer is confined onto the substrate by an LJ potential with
$\epsilon_{sa}=0.4\epsilon$, $\sigma_{sa}=1.2\sigma$, and $r_c=1.8\sigma_{sa}=2.16\sigma$.
The stronger binding energy $\epsilon_{sa} > \epsilon_{aa}$ gives perfect wetting and,
combined with the chain length, leads to a very low vapor pressure.
All simulations are performed with the substrate and adsorbate atoms 
kept at a constant temperature $T=0.175\epsilon/k_{\rm B}$.
This is lower than the melting point ($T=0.7\epsilon/k_{\rm B}$) of a LJ 
solid whose interactions are comparable to those in the substrate.
However, since $\epsilon_{aa}$ is only $0.25\epsilon$,
the temperature $T=0.175\epsilon/k_{\rm B}$ is higher than the
glass transition temperature ($T_g \sim 0.4\epsilon_{aa}/k_{\rm B} \sim 0.1\epsilon/k_{\rm B}$) 
in bulk systems~\cite{baljon96,rottler01}.
The large $\sigma_{sa}$ reduces the periodic potential from the substrate that might
lock the adsorbed film in an epitaxial state \cite{thompson90a}.
As a result, the adsorbed film is in a fluid state with a high mobility along the substrate surface.

The adsorbates also interact with the tip ($t$) through the LJ potential. 
The corresponding parameters are
$\epsilon_{ta}=0.75\epsilon$, $\sigma_{ta}=1.0\sigma$, and $r_c=2^{1/6}\sigma$.
These parameters ensure that the adsorbed film does not wet the tip. 
Thus the formation of a meniscus is avoided.

The simulations are performed using the Large-scale Atomic$/$Molecular Massively Parallel Simulator (LAMMPS) 
developed at Sandia National Laboratories. 
This classical MD code utilizes spatial decomposition to parallelize the computations.
Forces are calculated with the help of neighbor lists.
A velocity-Verlet algorithm with a time step $dt=0.005\tau$ is used to integrate the equations of motion.
Tests with $dt=0.001\tau$ to $0.007\tau$ give essentially the same results.
To fix the temperature $T$ of substrate and adsorbate atoms, a Langevin thermostat is applied
by adding a drag term to their equations of motion.
The Langevin damping rate $\Gamma$ is typically $0.5\tau^{-1}$.
To check that the thermostat did not affect the results, 
we also ran simulations with $\Gamma = 0.1\tau^{-1}$ and damping
only on directions perpendicular to the sliding velocity in friction measurements. 
No noticeable change was found.

Although nearest-neighbors in the substrate interact with ideal springs, 
changes in orientation of the springs lead to a slight anharmonicity. 
Instead of expanding with increasing $T$, 
the equilibrium lattice constant of the substrate shrinks by about $0.16\%$
as $T$ rises from $0$ to $0.175\epsilon/k_{\rm B}$.
The periodic boundary conditions prevent contraction in the $x-y$ plane, 
leading to a small negative pressure ($\sim -0.08\epsilon/\sigma^3$) in this plane. 
The height of the substrate shrinks about $0.3\%$ 
to relieve the tension along the $z$ direction. 
To check that this slight anisotropy did not affect our results, 
we also ran simulations in which the solid substrate was allowed to relax to
a fully equilibrated state with a zero inner pressure before the contact occurs.
No change was noticed for the results presented in this paper.

\begin{figure}[htb]
\centering
\includegraphics[width=3in]{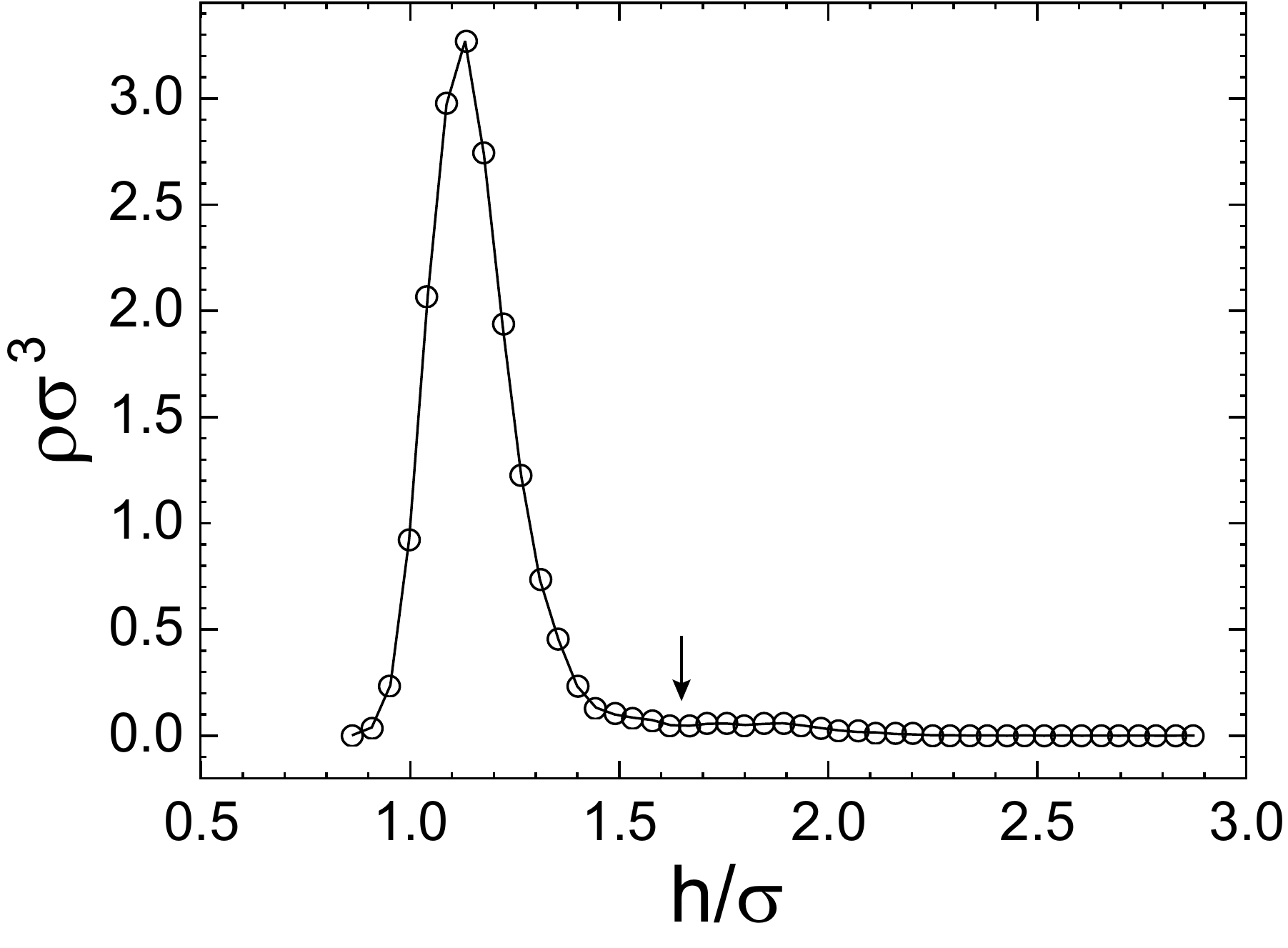}
\caption{Number density $\rho$ of adsorbate atoms vs. height $h$ above
the top layer of substrate atoms before the tip is brought into contact. 
Only points with a nonzero value are plotted. Lines are a guide to the eye.
The height 1.65$\sigma$ of the density minimum (arrow) after the first peak is
taken as the top of the fluid layer.}
\label{FluidVertical}
\end{figure}

A layer of adsorbate molecules is put on top of the substrate and
the system is allowed to relax into an equilibrium state at $T=0.175\epsilon/k_{\rm B}$.
In Fig.~\ref{FluidVertical} we plot the atomic number density $\rho$ of adsorbates as a function
of distance $h$ above the mean height of atoms in the top layer of the substrate.
The film is close to an ideal monolayer with a single major peak at about $h \sim 1.2\sigma$.
The adsorbate distribution has a long but weak tail at large $h$ and 
a small density peak corresponding to a second layer at $h\sim 1.9\sigma$.
Only $2\%$ of the atoms are outside the first layer 
and virtually no evaporation is observed.
However, the tail in density makes it difficult to define the separation
between the tip and monolayer covered substrate.
We will use the minimum in density at $1.65\sigma$ as a reference
height corresponding to the top of the fluid layer.

After the adsorbed film and the substrate are fully relaxed, a tip is brought into proximity.
The normal load on the tip is controlled and increased by small steps.
When the contact reaches a steady state at a given load,
the responses of the adsorbate and substrate are measured.
The local pressure is determined by finding all atoms on the tip or 
substrate that have a nonzero interaction with the monolayer or opposing surface. 
The normal component of the force on the atom is then 
divided by the area per atom to yield a pressure.
We also measure the density distribution of the adsorbates, the contact area, 
the tip displacement, and the substrate deformation.
Friction is measured by displacing the tip and monitoring the lateral force. 
The tip is moved at a constant velocity $v$ or attached to a constant velocity
stage with a spring of stiffness $k_s$.
The spring only affects the displacement along the pulling direction and
no motion is allowed along the perpendicular $y$ direction.

\section{Results of Molecular Dynamics Measurements}

The following subsections compare our simulation results to continuum solutions 
for the geometry studied in this paper, 
a sphere of radius $R$ pushed by a normal load $N$ into a flat elastic substrate.
Hertz theory for non-adhesive contacts predicts that the region of
contact is a circle whose radius $a$ is \cite{johnson85}
\begin{equation}
\label{HertzRadius}
a=(\frac{3NR}{4E^{*}})^{1/3}.
\end{equation}
The contact area is $\pi a^2$ and is thus proportional to $N^{2/3}$.
The pressure in the contact region takes the form
\begin{equation}
\label{HertzPressure}
p(r)=\frac{2aE^{*}}{\pi R}\sqrt{1-\frac{r^2}{a^2}},
\end{equation}
where $r$ is the distance from the center of the contact circle.
Hertz theory also predicts that the normal displacement $\delta$ after initial contact is given by
\begin{equation}
\label{HertzDeform}
\delta = a^2/R.
\end{equation}
Combining with Eq.~\ref{HertzRadius} this implies that $\delta$ increases as $N^{2/3}$.

\subsection{Distributions of Pressure and Adsorbed Molecules}

\begin{figure}[htb]
\centering
\includegraphics[width=3in]{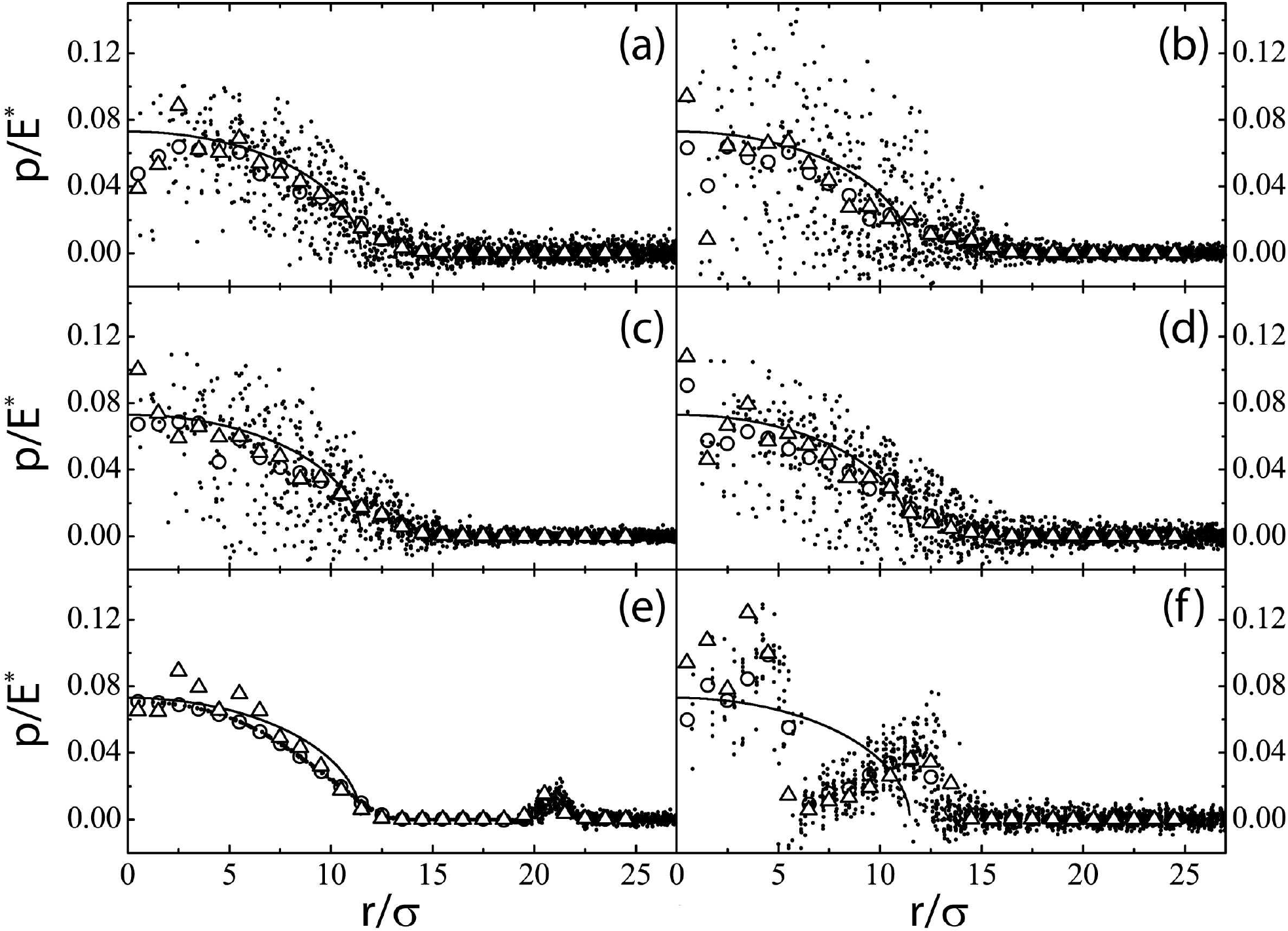}
\caption{Normal pressure distribution in the contact zone for different tip geometries:
(a) dense, (b) amorphous; (c) incommensurate; (d) commensurate out of registry; 
(e) commensurate in registry; (f) stepped. 
The pressure on each substrate atom is shown by a dot in the panels and 
the average over all atoms near a given radius is shown by circles. 
Only the average (triangles) is shown for the pressure on the tip atoms. 
Solid lines indicate the Hertz prediction, which is the same in all cases.
The normal load is relatively high: $N/(R^{2}E^{*})=2.01\times 10^{-3}$.}
\label{pressure}
\end{figure}

In Fig.~\ref{pressure} the normal pressure on the substrate surface is plotted 
as a function of radial distance $r$ from the center of the contact zone 
under a normal load $N/(R^{2}E^{*})=2.01\times 10^{-3}$.
The Hertz prediction using the bulk modulus of the substrate
is plotted as a solid line and is the same for all tips.
Small dots represent the raw data on every substrate atom 
and circles are the average over substrate atoms
in an annular region from $r-0.5\sigma$ to $r+0.5\sigma$.
Open triangles represent the angle-averaged normal pressure on the tip surface.
The forces on the substrate and tip must balance and from Saint-Venant's principal~\cite{love27} 
they are only expected to be redistributed by a distance of order the film thickness. 
The plots show that the two pressure distributions are very similar. 
The only pronounced deviations are near the center where
fewer atoms contribute to the average.
For example, only four atoms on the substrate surface are located within $r<1\sigma$.
As a result, the pressure distribution is more sensitive to the specific atomic arrangements 
at small radii and shows stronger fluctuations.

The raw pressure data exhibits very strong variations with angle.
These variations persist even after long averaging times, 
but are different for different initial conditions. 
Direct observation of the dynamics of the adsorbed molecules shows 
that those in the center of the contact region ($r/\sigma \lesssim 10$) 
are frozen in a glassy state. 
They undergo thermal oscillations that are too small to 
sample different pressure distributions in the contact
and very slow aging that is discussed further below. 
Similar glass transitions have been observed 
in a wide range of experiments~\cite{gee90,granick91,klein95} 
and simulations~\cite{thompson92,gao97c,cui03} on confined films. 
The fluid outside the contact remains in a mobile fluid state and 
the dynamics slows down dramatically at the outer edge of the contact. 

Previous studies of dry contacts with the same tips showed large changes 
in the distribution of contact pressure with tip geometry. 
Fig.~\ref{pressure} shows that tip geometry is less important 
when adsorbed molecules are introduced.
Tips with very distinct surface configurations 
lead to very similar pressure distributions (Figs.~\ref{pressure}a-d).
Structural effects are only evident for a bent commensurate tip 
that is in registry with the substrate surface (Fig.~\ref{pressure}e) and 
a tip cut from a crystalline solid (Fig.~\ref{pressure}f).
Moreover, these structural effects only become significant at high loads. 
When $N/(R^{2}E^{*}) \lesssim 2.5\times 10^{-4}$, all tips show similar $p(r)$.

\begin{figure}[htb]
\centering
\includegraphics[width=3in]{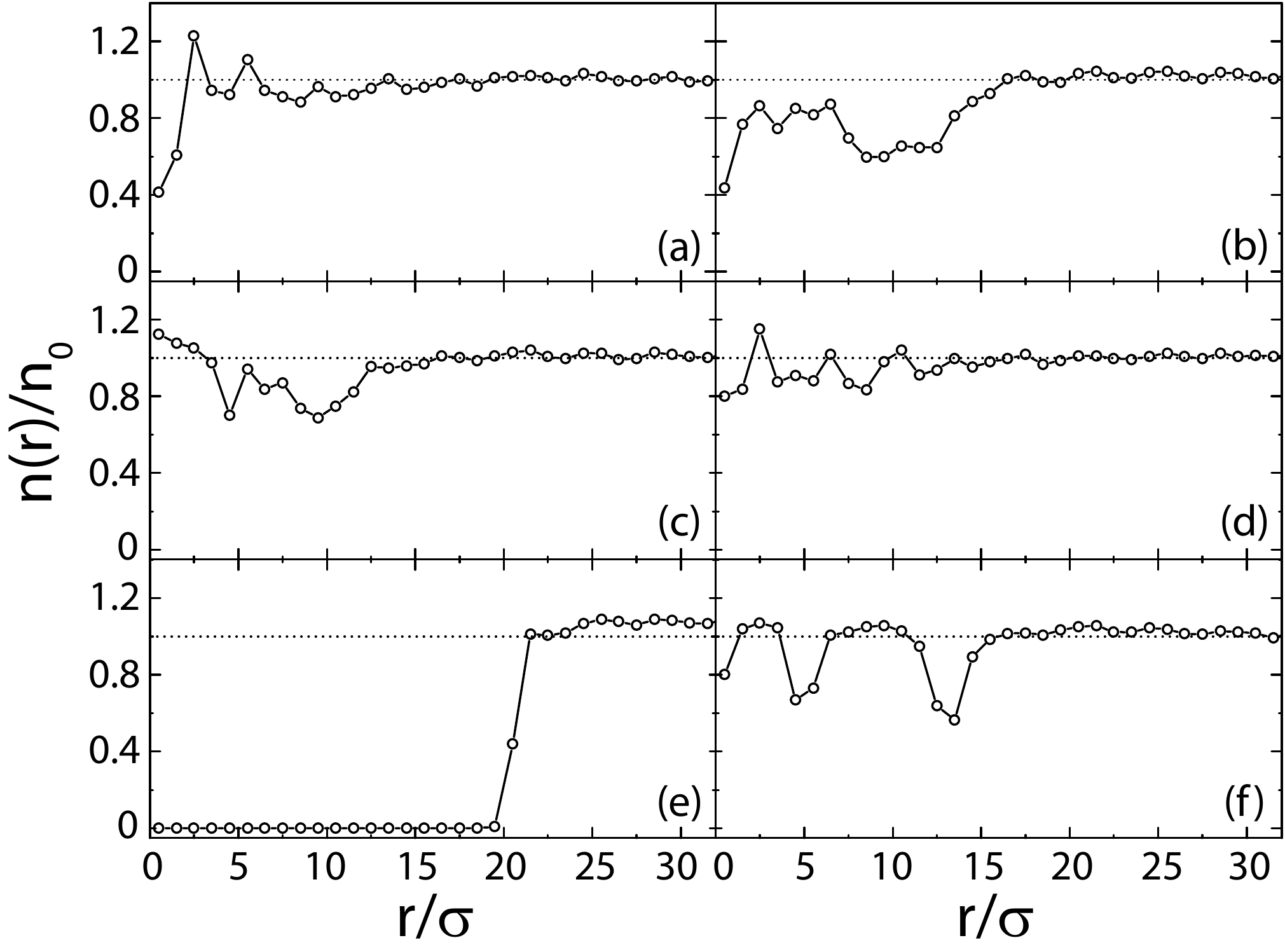}
\caption{Number of monomers per unit area $n(r)$ as a function of 
radial distance $r$ from the central axis ($r=0$) of the tip for different tip geometries: 
(a) dense, (b) amorphous; (c) incommensurate; (d) commensurate out of registry; 
(e) commensurate in registry; (f) stepped.
The normal load $N/(R^{2}E^{*})=2.01\times 10^{-3}$ is the same as for Fig.~\ref{pressure}
and $n(r)$ is normalized by the value before contact $n_0=0.794\sigma^{-2}$.
}
\label{FluidRadial}
\end{figure}
To examine how the pressure in the contact affects the adsorbed film, 
the adsorbate surface density $n(r)$ is plotted in Fig.~\ref{FluidRadial}. 
As in Fig.~\ref{pressure}, four of the tips have very similar density distributions 
(Fig.~\ref{FluidRadial}a-d) even though their different atomic structure led to 
very different behavior of bare tips~\cite{luan05,luan06b}. 
The density in the contact is reduced slightly and 
shows fluctuations that do not average out with time because the film is in a glassy state. 
The density decrease is slightly larger for the amorphous tip, 
perhaps because the greater surface roughness enhances mobility.

Figure \ref{FluidRadial}f shows two dips in $n(r)$ that correspond to 
the edges of terraces on the stepped tip. 
In continuum theory these edges produce a stress singularity, 
and substantial pressure rises are observed for bare tips. 
The adsorbed film smooths these peaks at terrace
edges into broad features in the pressure plot of Fig.~\ref{pressure}f.
Note that the sequence of terrace sizes is not uniquely determined
by the tip radius and this would lead to variations in the location
of these features~\cite{luan06b}.

Figure \ref{FluidRadial}e shows that the pressure from the commensurate tip in registry 
expels all adsorbed molecules from the central region. 
The tip atoms interact directly with the substrate, 
leading to a smooth pressure peak in the center of the contact (Fig.~\ref{pressure}e) 
that is similar to that for a bare surface~\cite{luan06b}.
A second pressure peak appears near $r=20\sigma$ 
where the tip contacts the edge of the adsorbed film.
For the in-registry tip, atoms from one surface are above gaps in the opposing surface. 
The space available to adsorbed molecules is fairly constant and 
the barriers to lateral motion are small. 
For the out-of-registry case, opposing solid atoms are 
directly above each other and adsorbed molecules can be 
trapped in pronounced channels between atoms. 
We found no expulsion of molecules for the out of registry tip even at the highest loads. 
In-registry tips expelled the film for $N/(R^{2}E^{*}) \gtrsim 2.5\times 10^{-4}$ and 
there was relatively little dependence on the rate of loading 
because the film was in a glassy state where thermal activation was negligible compared to
mechanical activation. 

All pressure distributions in Fig.~\ref{pressure} exhibit a long tail that extends beyond 
the contact region predicted by Hertz theory.
Two factors contribute to this. 
The first is that the adsorbed film is compliant and thus can conform to the tip,
allowing the pressure at the edge of the contact zone to drop to zero more smoothly 
than in the Hertz model. 
Note that the relatively small density drop for most tips indicates that this
happens mostly through compressing the layer rather than displacing atoms.
The second factor is thermal fluctuations in the substrate and adsorbed film. 
These are normally neglected in contact theories and are discussed in the next subsection. 

\subsection{Measuring Contact Area}

There has been great interest in the relation of single-asperity contact area 
to load and friction~\cite{carpick97,luan06a,luan06b,szlufarska09,knippenberg09,luan09,landman04}.
Fig.~\ref{pressure} shows that pressure is distributed over a larger area than the Hertz prediction.
However, the meaning of contact is somewhat ambiguous at the atomic scale.
Here we will consider three different definitions of contact radius and area.
If we define the contact radius as the distance from the contact center at which
the pressure on the tip surface vanishes (or drops below some threshold),
then Fig.~\ref{pressure} indicates that a value bigger than the Hertz prediction will be obtained.
We will call the contact radius associated with the edge of the pressure
distribution the outer radius $a_o$.
It is obtained from a linear fit to the tail of $p(r)$ on the tip surface. 
Systematic uncertainties are of order $\sigma$.

The outer radius is sensitive to the periphery 
of the contact where pressures are small. 
It is also interesting to consider a measure that is most sensitive to high pressure regions, 
such as the second moment of the pressure.
If the pressure distribution $p(r)$ in the contact zone follows 
the Hertz prediction (Eq.~\ref{HertzPressure}), 
then the contact radius $a$ satisfies
\begin{equation}
\label{secondmoment}
a^2=\frac{2}{5} \times \frac{\int_0^{\infty}r^2p(r){\rm d}^{2}r}{\int_0^{\infty}p(r){\rm d}^{2}r}.
\end{equation}
The right hand side of Eq.~\ref{secondmoment}
is readily calculated for the actual pressure distributions in our contacts.
This gives another measure of the contact radius that we will refer to as
$a_s$, where ``s" stands for second moment.
Greenwood and Tripp~\cite{greenwood67} considered a similar measure
in a study of the effect of roughness on the contact between
a sphere and a flat surface using the assumption that asperities could
be treated independently.
Their radius $r^*$ corresponded to the
ratio of the integral of the pressure divided by the integral of $p(r)/r$.
This measure is not plotted below because it is less common than
the second moment and
the fluctuations shown in Fig. \ref{pressure} make $r^*$ more sensitive
to noise than $r_s$.
However $r^*$ shows the same trends as $r_s$ and the deviations are
comparable to the numerical fluctuations.

Several groups have attempted to determine the contact area $A_c$ 
and the corresponding contact radius $a_c$
by counting the number of atoms $N_c$ on one surface that are within some
cutoff distance of the opposing surface~\cite{luan06a,szlufarska09,knippenberg09}.
However, this definition is very sensitive to the precise atomic structure of the tip
and substrate and also to the cutoff distance.
Often the cutoff is taken as the point where the potential becomes repulsive~\cite{burnham91}.
This criterion for contact correlated most simply with friction in a previous
study at zero temperature~\cite{luan06b} and corresponds to the cutoff in the
interfacial potential used here.

\begin{figure}[htb]
\centering
\includegraphics[width=3in]{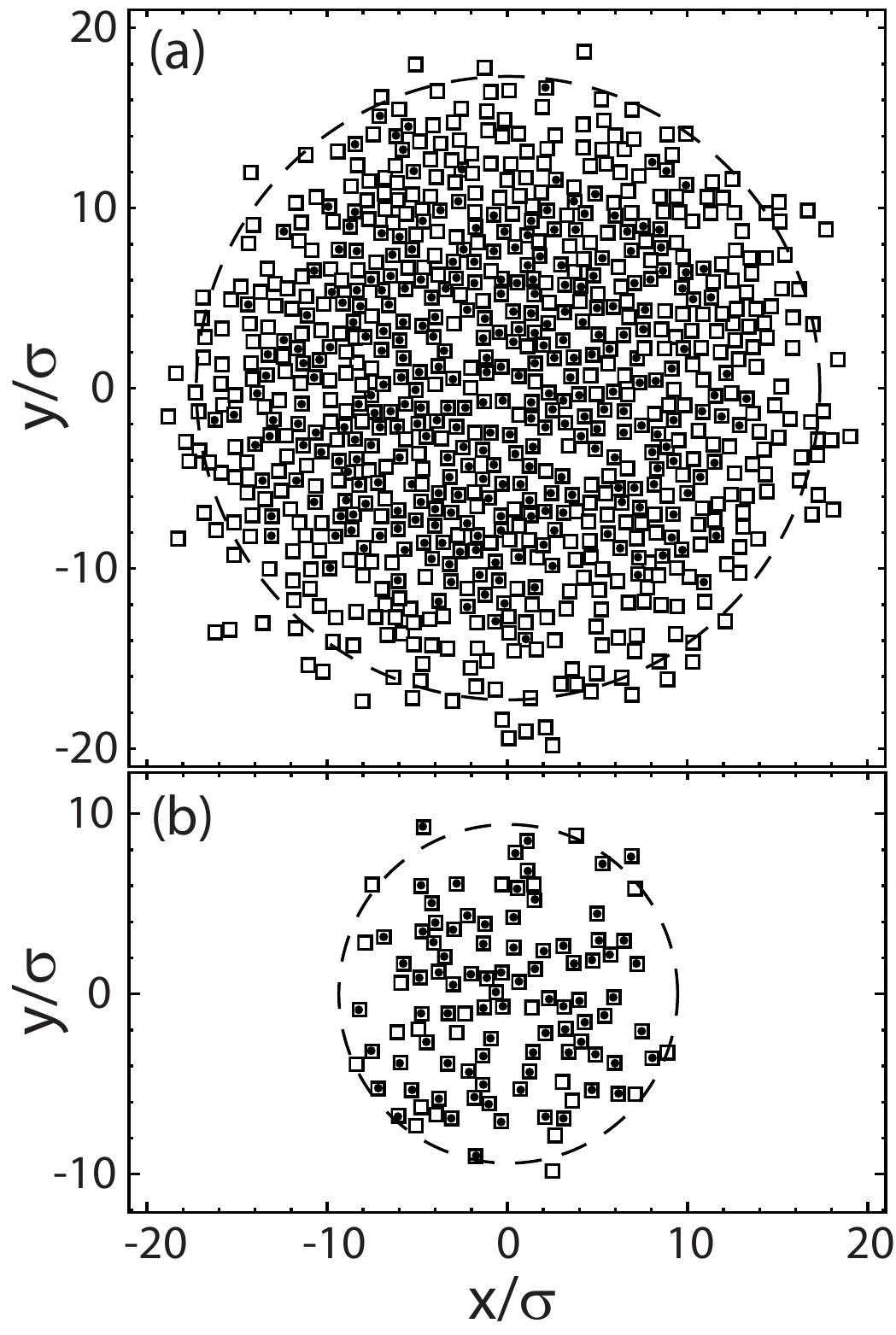}
\caption{
Atoms of an amorphous tip that contact over a time interval of
$50\tau$ ($\bullet$) or $2500\tau$ ($\square$) for a substrate
(a) with or (b) without an adsorbed film at $N=158 \epsilon/\sigma$.
Large dashed circles show the radius $a_o$ determined from the pressure
distribution.
This is insensitive to time interval.
}
\label{WetDry}
\end{figure}

In our current simulations we found that thermal fluctuations lead to
additional ambiguities in $N_c$.
Fig.~\ref{WetDry} shows all atoms on an amorphous tip that feel a repulsive
force in bare and wet contacts 
at $T=0.175k_{\rm B}/\epsilon$ and $N/(R^{2}E^{*})=2.5\times 10^{-4}$.
There is clearly a pronounced increase in the size of contact when an adsorbed
layer is present, but there is also a strong dependence on 
the observation time.
Different symbols indicate the atoms that contacted during 
time intervals of $50\tau$ and $2500\tau$.
The number of contacting atoms, and thus $A_c$ increases by about $25\%$ for
the bare contact and by a factor of $2.5$ for the wet contact.

\begin{figure}[htb]
\centering
\includegraphics[width=3in]{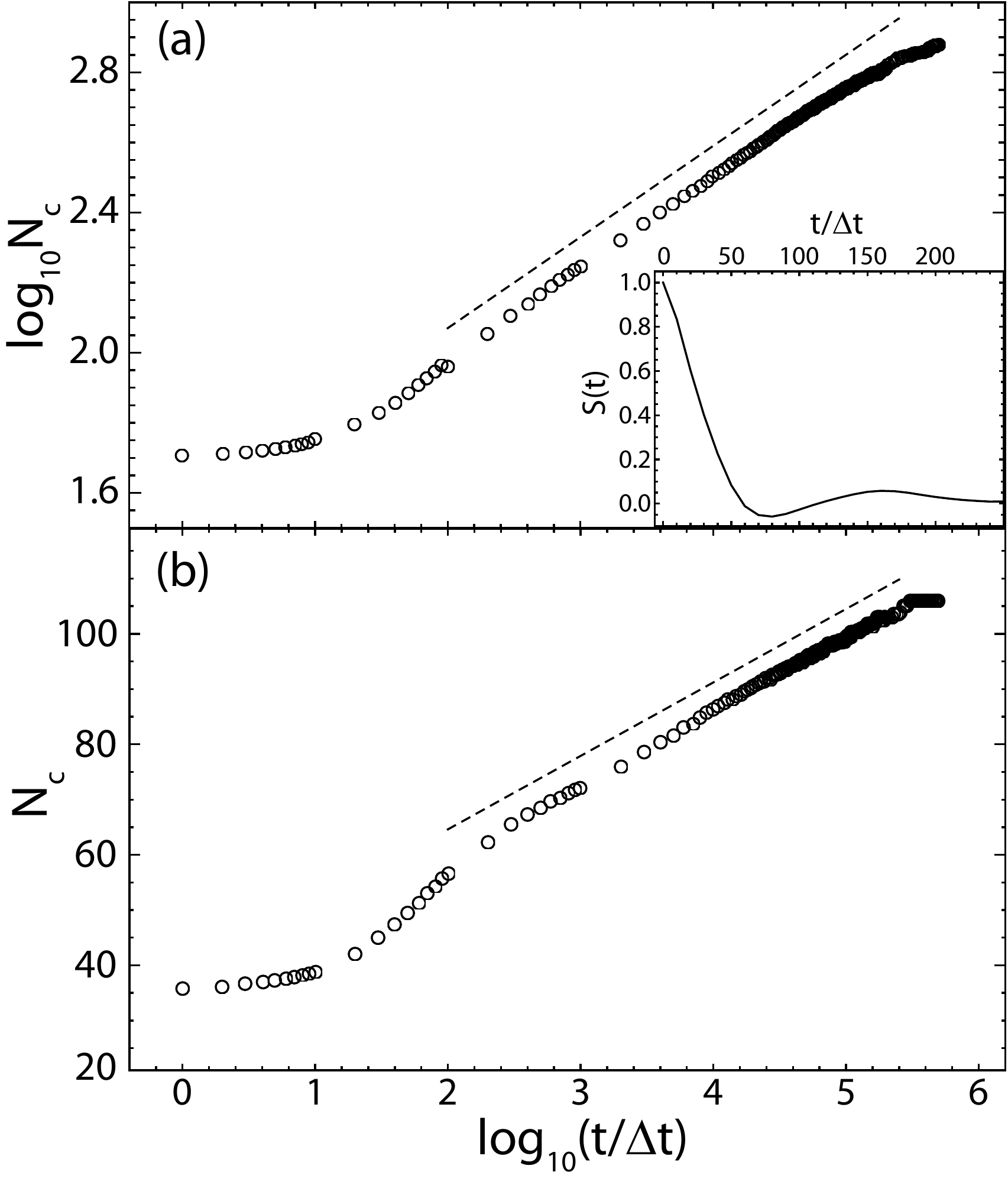}
\caption{The number of atoms $N_c$ on an amorphous tip
that contact the substrate or adsorbed layer over some part
of an interval containing $t/\Delta t$ MD time steps.
For both wet (a) and bare (b) contacts, the amorphous tip is used. 
Other tips produce similar results. 
For the wet contact, ${\rm Log}N_c \propto {\rm Log}t$ at large $t$, 
which means that $N_c$ has a power law dependence on $t$. 
The slope of the dashed line in (a) is $0.26$. 
For the bare contact, $N_c$ increases linearly with ${\rm log} t$ at large $t$. 
The slope of the dashed line in (b) is $13.3$.
The inset in (a) shows the autocorrelation function for an atom to remain
in contact.
}
\label{ContactAtoms}
\end{figure}

The variation of $N_c$ with the number of time steps $t/\Delta t$ is shown
in Fig.~\ref{ContactAtoms}.
Values of $N_c$ over a given time interval were averaged over many starting times 
once the system had reached steady state.
These results are for the amorphous tip, but similar results are obtained
for tips with other geometries.
For both bare and wet contacts
the number of atoms in contact is constant at short times and then increases monotonically.
The length of the initial plateau is about a quarter of the 
period of the most rapid vibrations
between neighboring atoms, i.e., the Einstein modes.
This is the minimum time for the configuration to change.
To confirm this we analyzed the autocorrelation function for contact
$S(t) \equiv \langle (\Theta(p(x,t'+t))-\langle \Theta(p(x)) \rangle)
(\Theta(p(x,t'))-\langle \Theta(p(x)) \rangle) \rangle$,
where $p$ is the local pressure, $\Theta$ the Heaviside function
and angle brackets indicate an average over $x$ and $t'$.
As shown for an adsorbed film in the inset of Fig.~\ref{ContactAtoms}a, 
$S(t)$ crosses zero after about $50$ time steps ($\sim 0.25 \tau$).
Deviations from the plateau in $N_c$ grow as $S(t)$ drops
and the asymptotic long time scaling of $N_c$ sets in at times
of order $100$ to $200$ $\Delta t$ when $S(t)$ is small.

The time dependence in $N_c$ comes from several factors.
For the bare contact the
dominant factor is thermal fluctuations of the substrate surface.
The longer the time interval, 
the more time for a rare event to bring atoms at large $r$ into contact.
The roughly logarithmic time dependence would be consistent with the
Boltzmann distribution and a linear increase in energy with the height
of surface atoms, but a proper calculation of $N_c (t)$ would
require a detailed analysis of normal mode energies and their effect
on contact.
For the wet contact, molecular rearrangements in the adsorbed film 
also contribute to changes in $N_c$, 
leading to more rapid growth with observation time:
$N_c \sim t^{\alpha}$ with $\alpha \sim 1/4$.
Slow diffusion and aging in the center of the contact brings different tip
atoms into contact.
At large $r$, the film is fluid and the cost for thermal fluctuations that
bring adsorbed molecules up into contact with
the tip is lower than for the bare contact.
This further broadens the region of contacting atoms.

In the following we consider the contact area associated with
the initial plateau in $N_c$, which 
is just the mean number of atoms in contact at any instant in time.
The corresponding contact area is then $A_c = N_c(\Delta t) A_{at}$
where $A_{at}$ is the area per tip atom.
This represents a lower bound on the area of contact because
of the monotonic rise in Fig.~\ref{ContactAtoms}.
It would be natural to define the area using a longer time interval
that averaged out the rapid oscillations due to high frequency phonon modes,
counting any atom that contacted over a period of $\sim 100 \Delta t$.
Even this short averaging interval increases the area by a surprisingly
large $50\%$ ($25\%$ increase in $a_c$).
This should be kept in mind when comparing to continuum theory, but
we do not present results for larger time intervals because it is hard
to identify any unique choice for the length of the interval.
If the contact region were circular,
the effective contact radius would be $a_c \equiv \sqrt{A_c/\pi}$ and 
this quantity is compared to Hertz theory.
Note that the contacting atoms are in general spread somewhat
dilutely over a significantly wider region, i.e., $a_c < a_o$.

Given the clear evolution in the number of contacting atoms with time interval,
one may wonder whether there is a corresponding evolution in the contact
radii measured from the pressure distribution.
The answer is no.
The radius determined from the second moment is particularly insensitive to time interval 
because it is dominated by the central regions of large pressure.
The value of $a_o$ increases only slightly with time, changing
by $0.5\sigma$ or less.
The large circles in Fig.~\ref{WetDry} show the values of $a_o$ determined from 
the long time interval.
Note that some atoms outside the circle were in contact during the interval.
However, the magnitude and duration of the forces during these contacts
were so small that they did not contribute significantly to the average
pressure and thus $a_o$.
Conversely, while many of the atoms inside $a_o$ do not contact during
the short time interval, we find that there are always a few atoms with
$r \approx a_o$ that make strong enough contacts to produce a significant pressure.
The identity of the atoms that make this contribution varies from one
time interval to the next, but the measured $a_o$ is nearly unchanged.

\subsection{Variation of Contact Area with Load}

\begin{figure}[htb]
\includegraphics[width=3in]{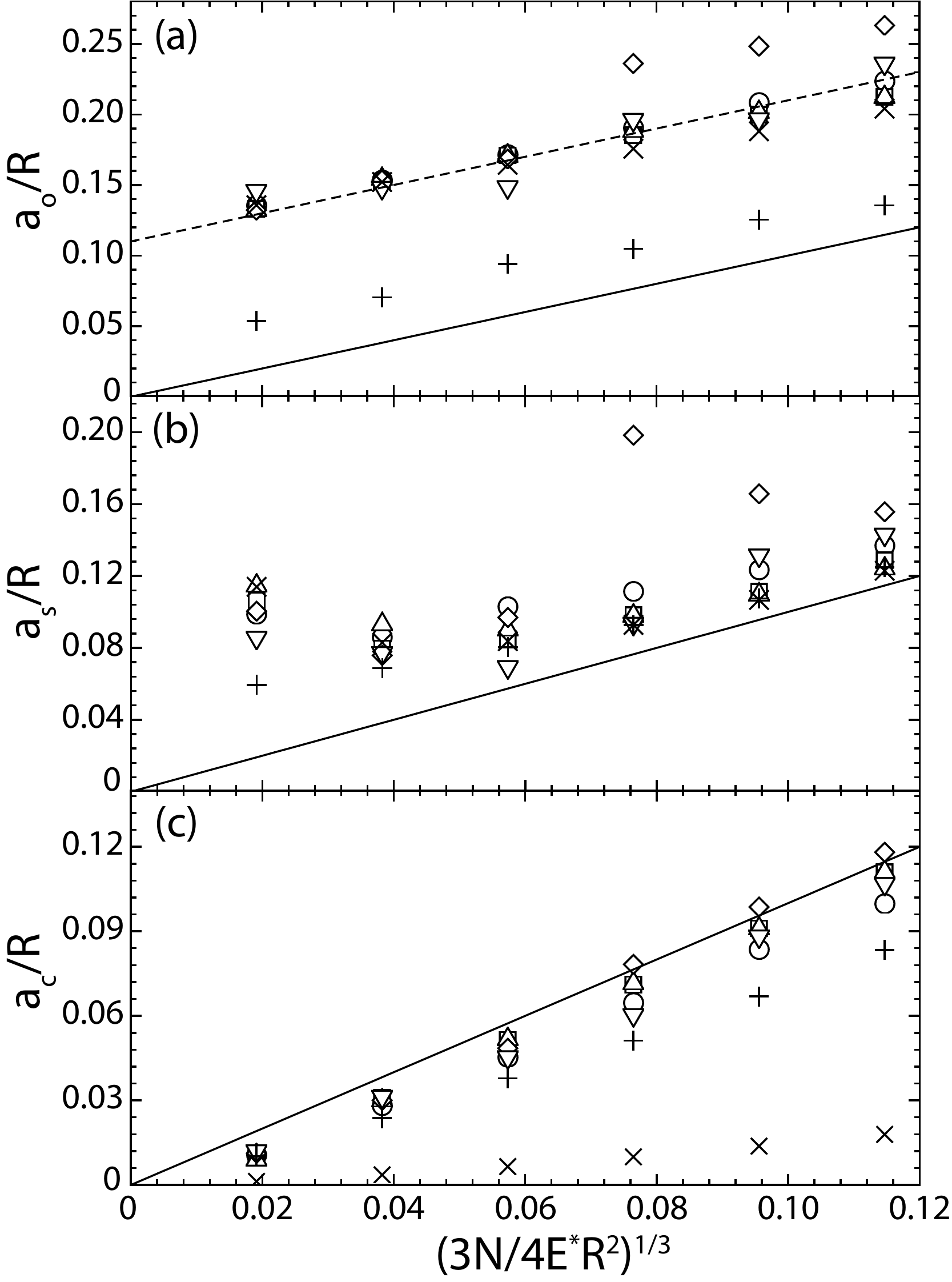}
\caption{Variation of contact radius $a$ with normal load $N$ for different tip geometries: 
dense ($\times$), amorphous ($\bigcirc$), incommensurate ($\Box$), 
commensurate out of registry ($\bigtriangleup$), commensurate in registry ($\Diamond$), 
and stepped ($\bigtriangledown$).
Results from Hertz theory (solid lines) and for a bare amorphous tip (+)
at the same temperature ($T=0.175 \epsilon/k_B$) are shown for comparison.
Three methods are used to measure the contact radius: (a) $a_o$ from the edge of $p(r)$; 
(b) $a_s$ from the second moment of $p(r)$; 
(c) $a_c$ from the mean number of contacting atoms at each MD time step.
}
\label{ContactRadius}
\end{figure}

Our results for the load dependence of different definitions
of the contact radius are shown in Fig.~\ref{ContactRadius}.
Values for the bare amorphous tip and the Hertz prediction (solid line)
are shown for comparison.
In most cases the radii rise linearly with $N^{1/3}$ and the slope
is close to the Hertz prediction.
However, there are large shifts between different measures of contact radius.

The radius $a_o$ where the pressure goes to zero (Fig.~\ref{ContactRadius}a)
shows the largest deviations from Hertz theory.
A significant increase over Hertz theory is observed for bare tips~\cite{luan05,luan06}.
Adding an adsorbed film spreads the pressure over a much larger area,
as already seen in Fig.~\ref{WetDry}.
Indeed the value of $a_o$ at the smallest load is bigger than the value
predicted by Hertz theory for the largest load we studied.
Note that at this radius, $a=0.14 R$, the tip surface is only
$0.01 R$ above the lowest point on the tip.
For $R=100\sigma$ this height is comparable to the width of
a molecular layer and it is not surprising that the film can
spread contact over this distance.
This effect is likely to be significant for most AFM tips, but
the broadening would be insignificant for the much larger radii
($R\sim 10$mm) in Surface Force Apparatus measurements.
The shift in $a_o$ for bare tips can also be attributed to a deviation from
hard sphere behavior.
However, instead of having an adsorbed layer that spreads interactions
over a height of order $\sigma$, the broadening comes from 
the shorter lengths associated with the range of repulsive interactions ($\sim 0.1\sigma$)
and amplitude of surface roughness~\cite{luan05,luan06}.

For the stepped tip, $a_o$ rises in discrete steps as $N$ increases. 
As shown in Fig.~\ref{ContactRadius}a,
it is almost constant for the lower three loads, 
jumps to another plateau for the next two loads, and has another discontinuous
increase at the largest load.
This is due to the fact that the surface of this tip has terraced steps.
Each time that the next terrace comes into contact with the adsorbed film, 
$a_o$ increases to the outer radius of that terrace.
Note that the sequence of terrace sizes is not uniquely determined
by the tip radius and thus the load where jumps
occur would vary with realization.

The pressure in the outer regions of the contact is small
(Fig.~\ref{FluidRadial}) and does not contribute
significantly to the load on the tip, particularly at large $N$.
Fig.~\ref{ContactRadius}b shows that the contact radius
determined from the second moment of the pressure is much
closer to the Hertz prediction.
The results at the largest loads are generally parallel to
the Hertz prediction and shifted upwards by only one or two
molecular diameters.
The only exception is the commensurate tip in registry.
As noted above, it expels the film at the highest three loads
and there is a second peak in the pressure at the edge
of the expelled film (Fig.~\ref{FluidRadial}e) that increases $a_s$.
The difference between bare and wet contacts for amorphous tips
is small, indicating that most of the contacting atoms in the outer
region of Fig.~\ref{WetDry}b carry relatively little load.
Similar results are obtained when comparing bare and wet
contacts of other tips.

At the lowest loads the value of $a_s$ saturates or actually increases
with decreasing load.
This counterintuitive behavior results from the very low pressures
in the contact.
The lowest load is $5.87\epsilon/\sigma$,
corresponding to a pressure of only $0.04\epsilon/\sigma^3$
when spread over the observed radius of order $7\sigma$.
The thermal contribution to the pressure in a system of density
$n$ is $n k_{\rm B} T$.
Thus the lowest load could be balanced by an ideal gas with
density of only $0.22\sigma^{-3}$
for the temperature in our simulations. 
Hertz theory ignores thermal fluctuations and can not
be expected to apply in this limit.
The pressures corresponding to the second and third loads
are about $8$ and $27$ times higher, respectively.
The thermal contribution to the stress becomes increasingly
irrelevant as the load increases and the Hertz prediction
becomes more accurate.
Note that at the lowest load we also find that the adsorbed
layer is not in a glassy state.

It is interesting to consider how the loads considered here
compare to typical experiments.
If we take $E^*=100{\rm GPa}$ and $R=30{\rm nm}$, the loads considered
here would range from about $1$ to $200{\rm nN}$.
This is comparable to the range of experimental loads.
However, experimental systems are often adhesive.
Even weak van der Waals interactions lead to forces per unit area
that are of order $k_{\rm B} T/\sigma^3$ or larger and this may
limit the role of thermal fluctuations.
Studies of this are underway.

Figure \ref{ContactRadius}c shows results for the radius $a_c$,
corresponding to the area of atoms in contact at any instant.
As for $a_s$, the increase in $a_c$ is roughly parallel to the Hertz
prediction, but $a_c$ is generally shifted to lower radii.
As noted above, $a_c$ is expected to be smaller than other measures.
Including all atoms that contacted on the time scale of high
frequency vibrations would increase $a_c$ by about $25\%$ and
make it comparable to $a_s$ at large loads for most tips.

The results for the dense tip indicate another difficulty that
arises in defining $A_c$.
Even for ideal flat surfaces, only a small fraction ($\sim 1/6$) of the atoms
on the dense tip are close enough to contact the opposing surface.
Atoms that lie between atoms on the opposing surface are too far away to contact.
Normalizing by the fraction of atoms that are in contact for two
flat surfaces gives values of $a_c$ that are quite close to
those for other tips.
Similar effects are likely to arise on surfaces with a range of surface
species, where some species may be screened by others with greater
height or larger interaction lengths. The relative contribution of
different species to contact area is also unclear in this case.

Note that the results on $a_c$ for the in-registry commensurate tip rise slightly above 
the Hertz prediction at the three largest loads.
This is because the tip breaks the adsorbed film and makes direct contact with
the substrate. If we only count the number of tip atoms in that direct
contact, then $a_c$ would be smaller than the Hertz prediction.
However, including contacts with the film at the edge of the contact zone 
gives a larger $a_c$.

\subsection{Normal Stiffness of the Contact}

Experiments can not measure the pressure distribution or contact 
radius directly, but the normal displacement can be determined
if the device is sufficiently stiff~\cite{kiely98}.
Hertz theory assumes hard sphere interactions and the tip
displacement is defined as the indentation $\delta$ relative to the first contact.
In experiments and simulations the interactions always have a finite
range that leads to ambiguity in the zero of $\delta$.
In our simulations, the long tail in the density 
shown in Fig.~\ref{FluidVertical}
leads to weak repulsions at large separations.
As noted in the discussion of this figure, we took
the density minimum at $1.65\sigma$ as the outer edge
of the first monolayer.
The tip would then contact this layer when height of the lowest tip atom
was higher than this outer edge by the cutoff radius $2^{1/6}\sigma$.
We use this height as our zero for $\delta$.

\begin{figure}[htb]
\centering
\includegraphics[width=3in]{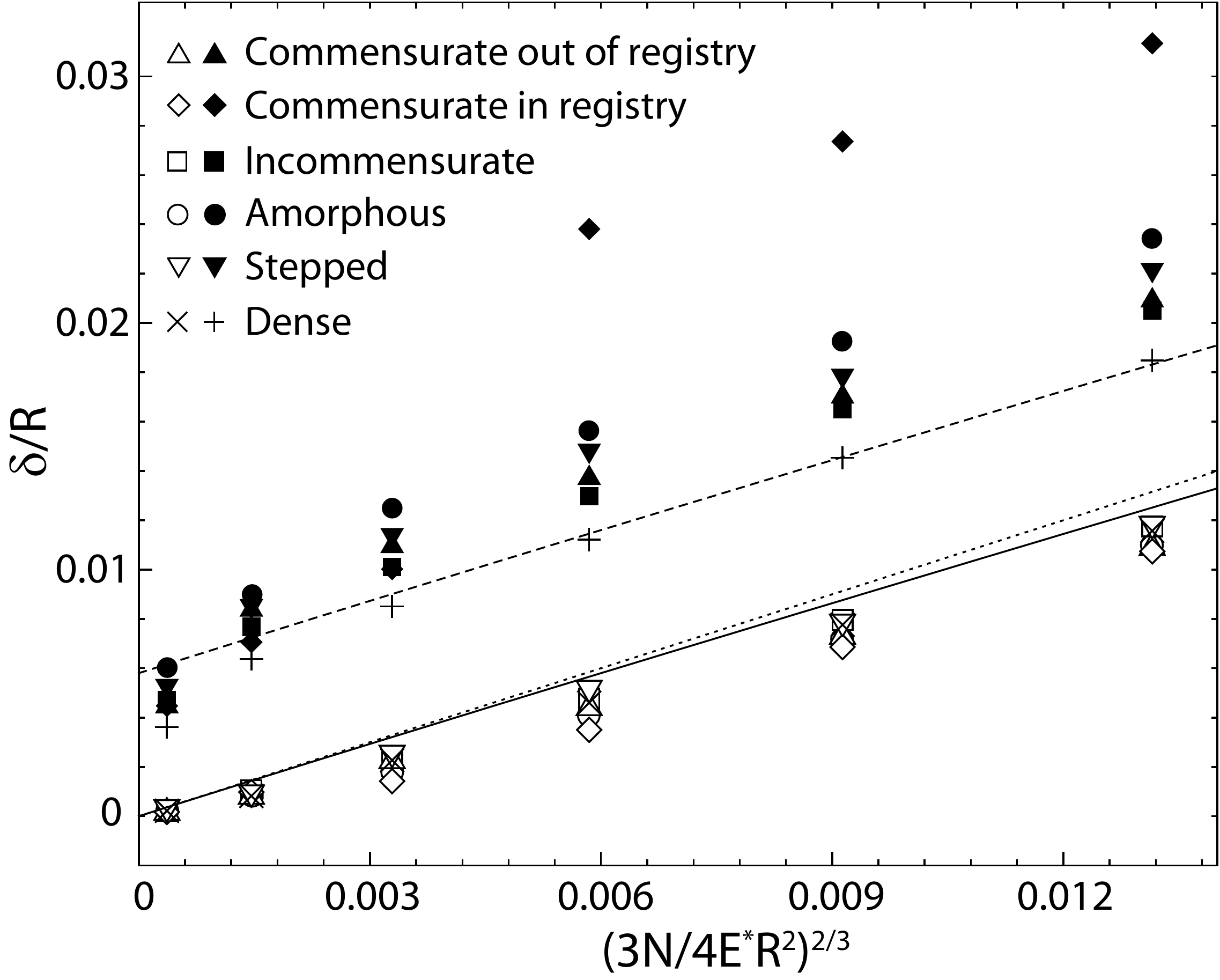}
\caption{The normal displacement of the tip $\delta$ (filled symbols) 
and of the substrate $\delta_s$ (open symbols) vs.
load for the indicated tip geometries.
The solid line is the Hertz prediction with $E^{*}=63\epsilon/\sigma^3$,
$R=100\sigma$ and a correction for finite substrate size~\cite{luan06b}.
The dotted line shows the uncorrected Hertz prediction and the
dashed line is the corrected result with a constant offset.
}
\label{deform}
\end{figure}

The filled symbols in Fig.~\ref{deform} show the normal displacement of the tip
$\delta$ for different tip geometries. 
The mean tip height was determined
by averaging over at least $2000 \tau$ at each fixed load.
Uncertainties due to thermal fluctuations were less than $0.05\sigma$,
and greatest at low loads.
The results are plotted against $N^{2/3}$ so that the Hertz prediction
for an infinite substrate is a straight line.
This is indicated by a dotted line.
A solid line shows the continuum prediction with a correction for
finite substrate depth.
The correction is less than $5\%$ and smaller than the symbol size.

For all tips, the value of $\delta$ rises much more rapidly than
predicted at low loads.
At higher loads, the rate of increase slows and results for most tips
can be fit by the Hertz prediction with a constant offset.
In experiments the modulus of a substrate is often
obtained from the slope of plots of $\delta$ against $N^{2/3}$.
Based on the results shown in Fig.~\ref{deform}, such fits should give
reasonably accurate moduli ($\sim 10\%$) if extended to sufficiently high load.

Similar load-displacement curves are obtained in calculations for substrates
that are covered by a thin layer that is more
compliant~\cite{sridhar04,adams06}.
When the contact radius is much smaller than the layer thickness,
indentation is described by the lower modulus of the layer
and $\delta$ rises more rapidly.
When the contact radius is much larger than the thickness,
the effective compliance is that of the stiffer substrate.
Plots like Fig.~\ref{deform} should show two straight lines
at low and high loads with slope changing by the ratio of $E^{*2/3}$.
The change of slope in our calculations by almost a factor
of three would imply an order of magnitude increase 
in modulus.
Indeed, as discussed above, the film may be more fluid than
solid at the lowest load.

To isolate the amount of displacement accommodated by the film,
we also evaluated the mean normal displacement of atoms in the top
layer of the substrate $\delta_s$.
Here the origin can unambiguously be taken as the mean height in
the absence of any load.
Moreover, Eq.~\ref{HertzDeform} corresponds exactly to the
expected substrate deformation for a rigid tip.

As shown in Fig.~\ref{deform}, $\delta_s$ is consistently below the
Hertz prediction for all tips.
The reason is that the adsorbed film distributes the force from
the tip over a wider area.
This increases the effective stiffness of the substrate and lowers
the displacement.
The broadening of the pressure becomes less important as the contact
grows and $\delta_s$ is parallel to the Hertz prediction at large loads.
The in-registry commensurate tip has a very large $\delta$ because
the adsorbed film is pushed out.
However, it produces a slightly smaller $\delta_s$ than those of other tips. 
This is because part of the load is spread to the fluid around
the tip and spreading the load over a larger area makes the
system stiffer.

The total change in $\delta$ is larger than that in $\delta_s$.
The difference must be accommodated by compressing the adsorbed film and the 
interfaces on either side.
For the cases studied this is typically less than $0.5\sigma$.
This total change should be insensitive to the radius of the tip.
Thus normalized plots like Fig.~\ref{deform} will show larger changes
for smaller $R$ and smaller changes for larger tip radii.

\subsection{Time Dependence of Friction}

Measurements of friction are known to be affected by the mechanical
properties of the measurement device, whether at macroscopic~\cite{bowden86}
or nanometer~\cite{muser03acp,medyanik06,conley08} scales.
To illustrate the range of possible friction forces, we contrast the friction
measured by stiff and compliant systems.
In the stiff case, the tip is displaced at a constant velocity $v$ along
the $x$ direction and the
only compliance is provided by the substrate and the interface with the tip.
In the compliant case, the tip is connected to one end of a spring of
stiffness $k_s$ and the other end is displaced at constant velocity.
The value of $k_s$ for each tip is chosen so that the spring is more
compliant than either the substrate or contact.
To limit the effects of inertia that are described below,
a damping force is added so that the spring is critically damped.
The damping does not contribute directly to the reported frictional forces.

\begin{figure}[htb]
\centering
\includegraphics[width=3in]{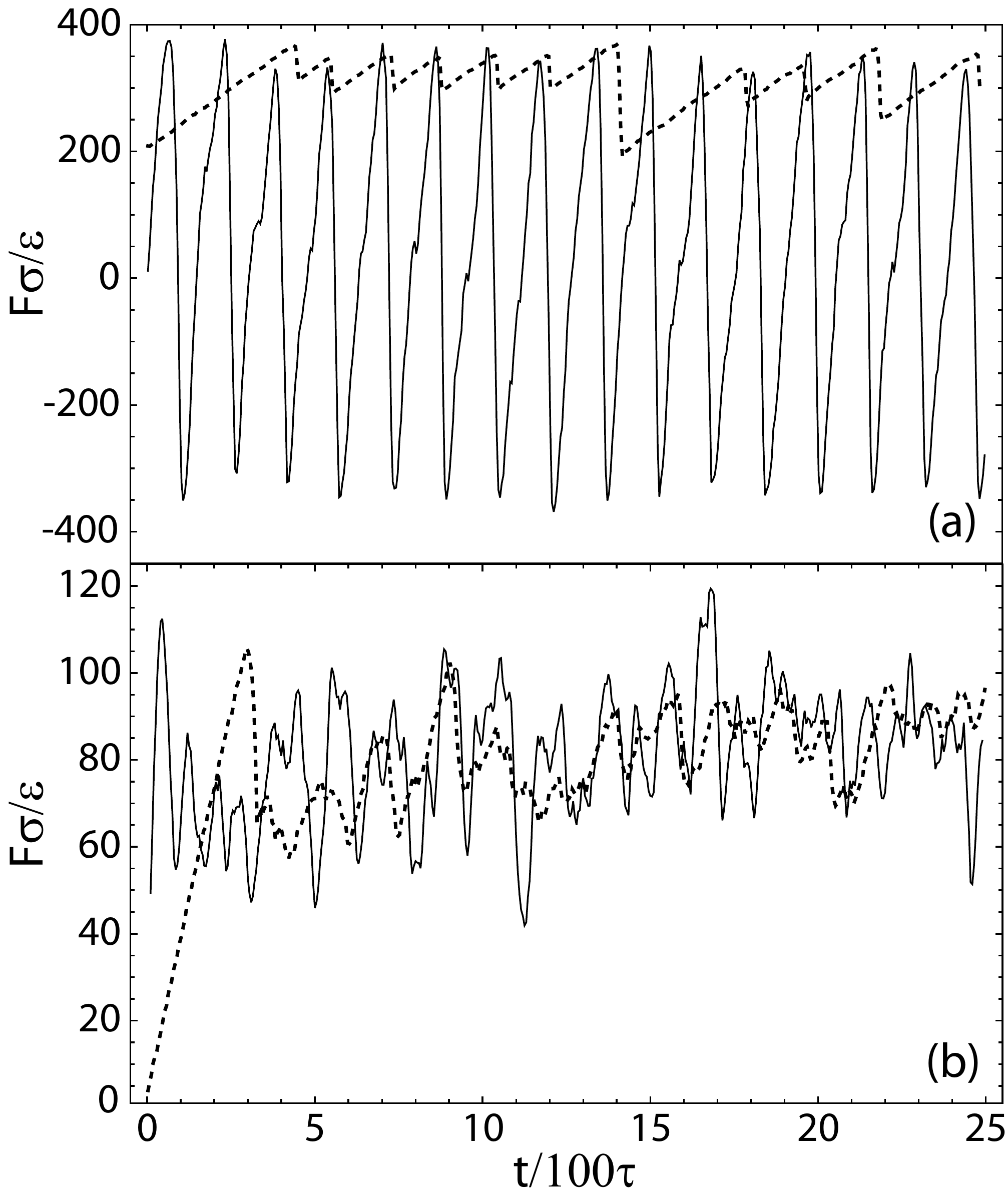}
\caption{Variation of friction force with time for
(a) out-of-registry commensurate and (b) amorphous tips driven
with constant tip velocity (solid lines) and through a compliant spring (dashed lines).
Since the velocity is $0.01\sigma/\tau$, a time interval of $150\tau$
corresponds to a displacement by the period $d=1.5\sigma$ of the
substrate along the direction of motion.
The normal load is $N=1267.2\epsilon/\sigma$ and 
the spring compliance is $k_s=40\epsilon/\sigma^2$.}
\label{FrictionTrace}
\end{figure}

Figure~\ref{FrictionTrace} shows the 
the time dependent friction
force on out-of-registry commensurate and amorphous tips
driven by stiff or compliant systems at $v=0.01\sigma/\tau$.
The commensurate tip exhibits simple, nearly periodic motion that is often
seen in experiments~\cite{carpick97}.
Due to the periodic structure of the tip, the system can find an equivalent
energy minimum after a displacement by a lattice constant $d$.
The amorphous tip is disordered and its motion is less regular.

The maximum friction force corresponds to the static friction $F_s$
that is needed to initiate steady sliding.
For both tips the heights of force maxima are relatively insensitive
to the compliance of the system.
However the heights do vary with position, particularly for the
amorphous tip.
These fluctuations reflect changes in the conformation of the adsorbed layer
and tend to increase in magnitude with velocity~\cite{luan04}.
The kinetic friction $F_k$ corresponds to the time average of the
friction force and varies strongly with compliance.
Stiffer systems have a lower $F_k$ because the
friction minima decrease with increasing stiffness
and may even become negative.

To understand these trends it is useful to consider the simple
Prandtl-Tomlinson model~\cite{prandtl28,tomlinson29,muser03acp}
for motion of a single degree of freedom,
the rigid tip, in a potential with period $d$.
The equation of motion is
\begin{equation}
m \ddot{x} = k_s (x_0-x) -F_0 \sin\frac{2\pi x}{d} - \Gamma \dot{x} \ ,
\label{Tomlinson}
\end{equation}
where $m$ is the mass, $x$ is the location of the tip,
dots indicate time derivatives,
$k_s$ is the stiffness of a pulling spring whose
other end is at position $x_0$,
and $\Gamma$ is a damping coefficient.
The static friction is just the
maximum force exerted by the periodic potential, $F_0$.
The kinetic friction exhibits two different
types of behavior depending on whether $k_s$ is smaller
or larger than a critical value $k_c \equiv 2 \pi F_0/d$.

For $k_s > k_c$ there is a single static solution to Eq.~\ref{Tomlinson}
for each $x_0$.
As $x_0=vt$ increases, the tip moves smoothly with $\dot{x} \approx v$
and the friction from the
periodic potential varies from positive to negative values.
The resulting time average force is small and $F_k$ goes
to zero with decreasing $v$.
This type of behavior is exhibited by the commensurate tip when
the only compliance comes from the substrate.
Note that the negative minima in the force are nearly equal and opposite
to the positive maxima.
The time averaged kinetic friction $F_k=34\epsilon/\sigma$ is
less than $10\%$ of
the static friction $F_s=368\epsilon/\sigma$.
To confirm that $k_s >k_c$ for this system, we note that
the values of $F_s$ and $d$ imply $k_c=1462\epsilon/\sigma^2$.
The compliance of the substrate is comparable to the prediction of
continuum theory $k_s=8 G a$, 
where $G=18.3 \epsilon/\sigma^3$
is the shear modulus and $a$ the contact radius.
Since all measures of $a$ discussed above are larger than the
continuum prediction of $11.5\sigma$
for this load, $k_s > 1684\epsilon/\sigma^2 > k_c$.
The same condition holds for all other loads because
$F_s$ decreases more rapidly with load than the contact radius does.

When $k_s < k_c$ there are multiple static solutions to Eq.~\ref{Tomlinson}
for a given $x_0$.
The tip remains stuck in one until it becomes unstable and then jumps
rapidly to another static solution.
The force is near $F_s$ before the jump, and only drops by of order
$k_s d$ since the distance between
stable states is of order the period.
As a result, the time average kinetic friction approaches $F_s$ as
$k_s \rightarrow 0$.
For the commensurate tip with a compliant spring,
$k_s = 40\epsilon/\sigma^2 << k_c$ and $F_k$ is almost $90\%$
of $F_s$.
Studies with weaker and stronger springs confirmed that
$F_k/F_s$ varies from nearly zero at $k_s > k_c$ to 
almost unity in the limit $k_s << k_c$.

The time dependence of the compliant system illustrates another effect
that can change the measured kinetic friction.
In Fig.~\ref{FrictionTrace}(a) 
most of the drops in force have the same size, but two later
drops are larger by a factor of three ($t \sim 1400\tau$) 
and two ($t \sim 2200\tau$), respectively.
In these cases the inertia of the tip allowed it to jump forward by more
than one period.
These multiple slip events are also observed in some AFM
experiments and have been modeled using Eq.~\ref{Tomlinson}~\cite{medyanik06,conley08}.
They lead to a substantial reduction in $F_k$ by reducing the
minima in the force.

The frequency and magnitude of multiple slip events vary with
load as well as compliance and tip inertia~\cite{medyanik06,conley08}.
While this may be important in certain experimental systems,
it makes it difficult to report a single number for the load
dependent kinetic friction on our model tips.
To minimize the associated ambiguity, we will focus on cases
where multiple slips have been eliminated by critical damping
of the spring and use of a sufficiently stiff pulling spring
for each tip.

The time dependence of the friction force on the stepped and
incommensurate tips is qualitatively similar to that of the commensurate tip.
In both cases, the tip can find an equivalent minimum if it moves
forward by a lattice constant of the tip while the adsorbed
film remains unchanged.
Note that translating the incommensurate tip produces a different
registry with the substrate, but this does not matter if the
adsorbed film has already adapted to the periodicity of the tip.

In contrast, the
time dependent friction on the amorphous tip is not simply described
by Eq.~\ref{Tomlinson}.
Because the structure of the tip is not periodic, the tip does not
find an equivalent potential energy minimum after moving by any distance.
Periodic motion would be possible if the adsorbed molecules slid rigidly
over the substrate with the tip, but this is not observed.
At low loads, sliding is localized at the interface between the tip
and film. 
The film is damaged at higher loads, as discussed in the next subsection.
While the tip drags some adsorbed molecules over the substrate, they
do not translate rigidly with the tip and thus do not lock into an
equivalent energy minimum after moving by a period.

The lack of periodicity has a direct impact on the friction traces
for amorphous tips.
Even for the stiff system,
where $k_s > k_c=420\epsilon/\sigma^2 $, the friction force remains positive
as the tip jumps forward.
As a result, $F_k$ is close to $F_s$ for any degree of compliance.
The minima in friction are typically about half the maxima rather than
being equal and opposite in magnitude as predicted by Eq.~\ref{Tomlinson}.
The negative forces for the stiff commensurate system come from the
intervals where the tip is being pulled forward as it drops into
a free energy minimum.
There is no preformed minimum for the amorphous tip to drop into
and thus no negative force.

For the stiff amorphous system the force shows oscillations
that are not periodic but have a typical spacing that is
of order the atomic diameter. 
This length scale enters simply because
it is the typical distance that
the tip must move to interact with different molecules after 
leaving a low free energy state.
For the compliant system the motion is even less regular.
There are some intervals where the
force is fairly constant, and the tip moves over the substrate
at a nearly constant velocity.
At other times the tip gets stuck in a local free energy minimum.
The force then rises linearly with time until the minimum becomes
unstable and the tip jumps forward.
Similar erratic stick-slip has been observed in previous simulations
and was found to be sensitive to the inertia of the tip~\cite{luan04}.
The simulations reported here are overdamped, which decreases the role
of inertia.

\subsection{Load Dependence of Friction}

The static and kinetic friction were obtained from friction traces like
those shown in Fig.~\ref{FrictionTrace}.
The kinetic friction is just the time average of the force after the
system reaches a steady state.
The static friction was obtained by identifying the peak forces during 
each interval where the surfaces locked together.
We found the peaks were easiest to identify for the compliant systems
and present those results below.
Values obtained from stiff systems were the same within the statistical
uncertainty, which was less than $10\%$.
Experiments, previous simulations, 
and simple models suggest that the static friction
should grow with the time the surfaces are stuck together
because of aging in the adsorbed film~\cite{dieterich79,ruina83,rottler05age,he01tl}.
This dependence is typically logarithmic in contact time and
difficult to separate from statistical fluctuations.
It is not considered here.

The static and kinetic friction were evaluated over a sliding distance
of at least $20\sigma$.
At low loads
the friction varied little over this interval.
At the highest loads some tips induced transitions in the state of the film
with sliding distance that changed the friction force.

\begin{figure}[htb]
\centering
\includegraphics[width=3in]{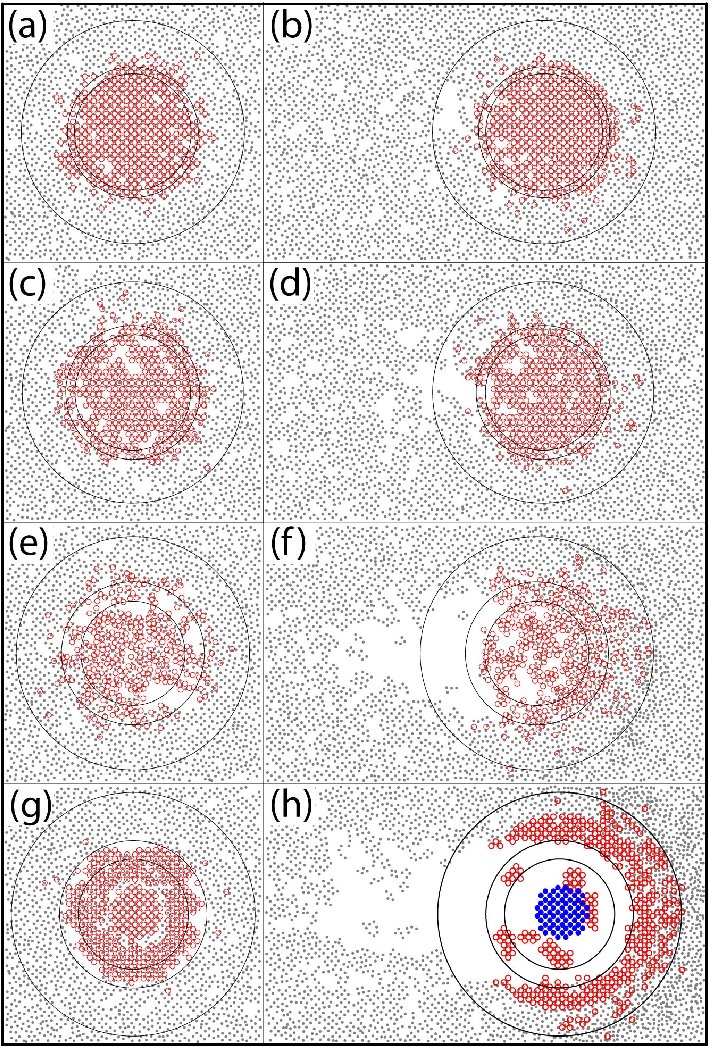}
\caption{(color online)
Snapshots showing the location of all film atoms (dots) and 
those tip atoms that contact the film (open red circles) or substrate
(filled blue circles) over an interval of $0.5\tau$
before (left) and after (right) the tip is pulled over the substrate.
The tip geometries are:
out-of-registry commensurate (a and b); incommensurate (c and d); 
amorphous (e and f); and stepped (g and h).
The vertical height of the panels is $50\sigma$ and the horizontal
dimension is the same for the left panels and $85\sigma$ for the right panels.
Large circles show the contact radii determined in Fig.~\ref{ContactRadius}
with the largest corresponding to $a_o$ and the smallest to $a_c$. 
The only place where tip atoms directly contact the substrate is within the
inner circle in panel (g).
}
\label{FilmDamage}
\end{figure}

The evolution of film structure at $N=1267.2\epsilon/\sigma$ is
illustrated in Fig.~\ref{FilmDamage}.
At this load, there is a tendency for a hole to form
behind the tip and for molecules to pile up in front of the tip.
The size of these effects increases with the roughness of the tip.
The smallest changes are for the bent incommensurate and out-of-registry commensurate tips.
Both slide on top of the film at all loads, with almost all of the
sliding velocity accommodated at the tip/film interface.
In contrast, the stepped tip completely expels film molecules from under
the first terrace for $N \gtrsim 375 \epsilon/\sigma$.
Molecules remain trapped under the second terrace and tend to move with the tip.
For $N=1267.2\epsilon/\sigma$, even the adsorbed molecules under the second terrace 
start to be squeezed out.
This allows the first terrace to contact the substrate directly
for the first time and leads to a rapid rise in friction. 

The amorphous tip slides on top of the film at low loads, but the mode
of sliding changes for $N \gtrsim 733 \epsilon/\sigma$.
Some of the molecules become trapped in the roughness of the tip and are
dragged along with the tip.
The sliding velocity is accommodated between these molecules and the substrate.
Because the molecules are trapped in positions related to the 
roughness of the disordered tip, they do not lock well with the
substrate and the friction rises less rapidly with $N$ than at lower loads.
The tip and entrapped film molecules push the rest of the film aside.
Molecules pile up in front and a bare patch is left behind.
This bare patch fills in fairly rapidly and the region where the
tip started has nearly healed by the end of the run.
Such healing of the lubricant layer could be very important to function
as a boundary lubricant.

A different type of 
time dependence was observed for the in-registry
commensurate tip at high loads ($N \gtrsim 375 \epsilon/\sigma$) and velocities.
For $v \gtrsim 0.01\sigma/\tau$ the tip starts in direct contact with
the substrate and the first friction peak is about
twice as large as for the out-of-registry tip.
Once sliding starts, the tip lifts up onto the film.
The friction drops sharply and then rises gradually,
with the final friction being very close to that for the out-of-registry
commensurate tip.
For this reason we do not show separate friction results for the
in-registry commensurate tip.
When the velocity was decreased to $v=0.002\sigma/\tau$
the tip remained in contact with the substrate during the sliding.
In this case the tip plows through the film and the force peaks
remain roughly twice the height of those at $v=0.01 \sigma/\tau$.

For both the in and out of registry bent commensurate tips, the sliding path
takes atoms on the tip directly over atoms in the substrate.
This is why the frictional forces are identical as long as the film
is either present or absent for both tips.
Previous work has shown that the friction on commensurate
tips changes as the line of motion is displaced perpendicularly to the sliding
direction~\cite{harrison93,perry97,harrison07}.
To illustrate the effect of perpendicular shifts, 
the stepped tip was displaced diagonally
by $0.5 d$ along $x$ and $y$ so the tip atoms
move along a line that is centered between substrate atoms.
Before displacing the friction on the stepped tip is close to that on bent tips.
After displacing the friction is reduced by a factor of about two,
with the precise ratio depending on load. 
Similar reductions are observed when the bent commensurate tip 
is offset in the same way.
The stepped tip was also made incommensurate by rotating
it around the $z$-axis by $20^\circ$.
This produced an even larger drop in friction.

\begin{figure}[htb]
\centering
\includegraphics[width=3in]{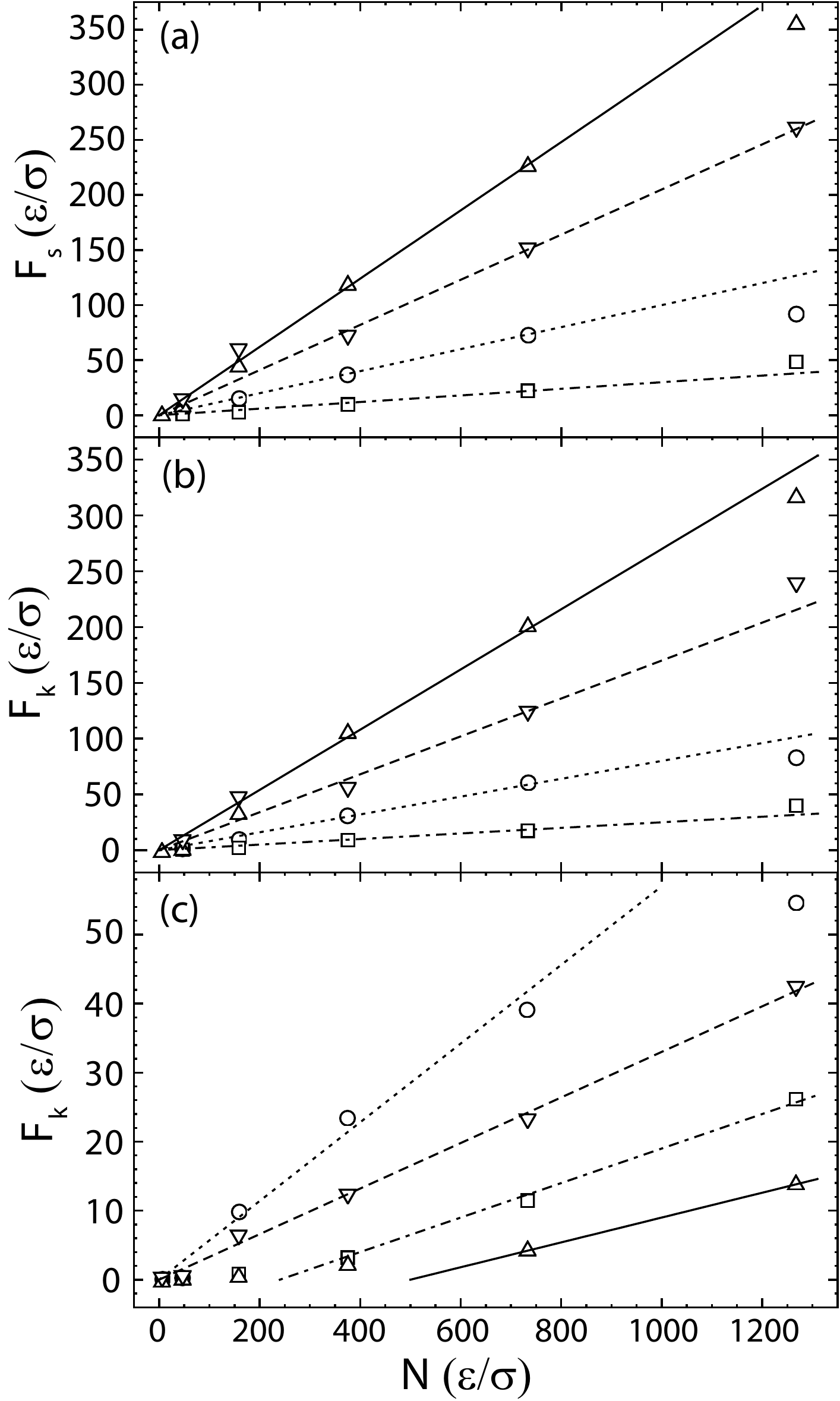}
\caption{
Static ($F_s$) and kinetic ($F_k$) friction vs. load ($N$) for the indicated tip geometries:
amorphous ($\bigcirc$), incommensurate ($\square$), 
out-of-registry commensurate ($\bigtriangleup$), and stepped ($\bigtriangledown$).
The static friction in (a) and the kinetic friction in (b) are measured by 
dragging the tip along the $x$-direction with a spring at a constant velocity $0.01\sigma/\tau$.
The kinetic friction in (c) is measured by displacing the tip uniformly 
along the $x$-direction at a velocity $0.002\sigma/\tau$.
Straight lines are linear fits to the friction.
Except in (c), they go through the origin.}
\label{fric}
\end{figure}

Figure \ref{fric} shows the load dependence of the static friction ($F_s$) on different tips
and the kinetic friction ($F_k$) obtained for compliant and stiff systems.
While the friction varies with tip geometry,
the deviations are much smaller than those observed for bare tips.
For the same tips, the friction in bare contacts varied by two orders of
magnitude and showed different functional dependencies on load.
Inserting the adsorbed layer has greatly reduced this sensitivity to tip geometry,
and the variation is even smaller if the special case of commensurate
tips moved directly over substrate atoms is removed.
In addition, the tips all exhibit a roughly linear rise in friction
with load, although there are transitions in behavior for some tip geometries.
For example the friction on the stepped tip increases more rapidly
when the film is expelled from the central terrace
($N \gtrsim 375\epsilon/\sigma$),
while the friction on the amorphous tip increases less rapidly 
with load when the film is damaged ($N \gtrsim 733\epsilon/\sigma$).

In these and previous simulations the largest static friction is observed
for the commensurate case where
the opposing surfaces have the same period and orientation.
This case also typically yields a linear relation between $F_s$ and load.
The simplest explanation is that the surfaces are effectively in hard-sphere
contact and the tip must be lifted up over atoms in the layer below
before it can slide forward~\cite{muser01prl}.
The friction provides the force needed to move the tip up
the ramp formed by these atoms
and the slope of the ramp gives $dF_s/dN$.
The slope is largest for the commensurate case because the effect
of all atoms adds coherently.
The slope is also larger when the tip atoms pass directly over the
substrate atoms rather than passing between them as for the
stepped tip simulations.
For incommensurate and amorphous surfaces, the forces average to zero
for bare, infinite surfaces.
This cancellation is prevented by the adsorbed film because it
adopts a glassy state that adapts to the structure of both tip and 
substrate~\cite{he99}.
However, the commensurate case still proves most easily locked together.

The kinetic friction for compliant systems is very similar to $F_s$.
The stiff amorphous results are also comparable to $F_s$ for the
reasons discussed in the previous section.
The friction force on commensurate systems is substantially reduced
in the stiff system.
The friction is nearly zero at small loads
and then rises linearly at large loads.
This behavior
can be understood from the Prandtl-Tomlinson model.
At low loads, $k_s > k_c$ and the kinetic friction vanishes with decreasing
velocity.
As the load increases, the potential and $k_c$ increase.
At sufficiently large load, $k_s$ may be smaller than $k_c$ leading
to a kinetic friction that follows the linear rise in $F_s$.
As expected from Fig.~\ref{FrictionTrace},
the stiff commensurate system has $k_s > k_c$ for all loads.
Even though the commensurate tip has the largest static
friction of any tip,
$F_k$ is the smallest and goes to zero with decreasing velocity. 

There has been great interest in understanding the connection between
the friction in molecular scale asperities and macroscopic measurements
of friction.
Amontons's laws state that macroscopic friction
is proportional to load and independent of the nominal area of 
the contacting surfaces.
However, macroscopic surfaces are generally rough and, as noted
above, the real area of molecular contact may be much smaller
than the nominal area.
In many cases, $A_{real}$ is expected to be proportional to load but
is difficult to measure, making it
unclear whether the area or load is controlling the friction force.
There are also many exceptions to Amontons's laws.
For example, adhesion between surfaces leads to friction at zero
or negative loads and grows in importance as dimensions shrink.
Repulsive tip interactions were used in our simulations to
eliminate adhesive effects.

The variation of friction with load in Fig.~\ref{fric} is roughly linear
and thus consistent with Amontons's laws.
As noted above, deviations from linearity tend to coincide with changes
in the structure of film and location of sliding.
Linear fits to the kinetic friction reach zero friction at a positive load,
but this shift depends on the pulling spring and can be understood from
the Prandtl-Tomlinson model (Eq.~\ref{Tomlinson}).

Studies of bare tips by Luan and Robbins showed that the friction did not
always increase linearly with load over the same range of loads studied
here~\cite{luan06b}.
In particular, the static friction on the amorphous and incommensurate tips
increased sublinearly and the stepped tip showed discontinuous jumps in friction.
When a film is present, the friction on the amorphous tip only rises
sublinearly at the highest load and
this is directly attributable to a change in sliding mechanism that
one would not expect to describe simply in terms of the change in load or area.

Mo et al. have recently considered the friction between
diamond and a hydrogen-terminated amorphous carbon tip,
using realistic interaction potentials~\cite{szlufarska09}.
They concluded that both the friction and the area obtained
from counting atoms, $A_c$, were proportional to load and contrasted
this with the results of Luan and Robbins.
The plots shown by Mo et al. actually show that neither friction nor area
is strictly proportional to load.
Rather both are zero up to a finite load and then rise linearly.
While this linear relation is still inconsistent with the sublinear
relation obtained by Luan and Robbins, the latter authors reported
the static friction while Mo et al. reported kinetic friction 
and went to much smaller dimensionless loads.

\begin{figure}[htb]
\centering
\includegraphics[width=3in]{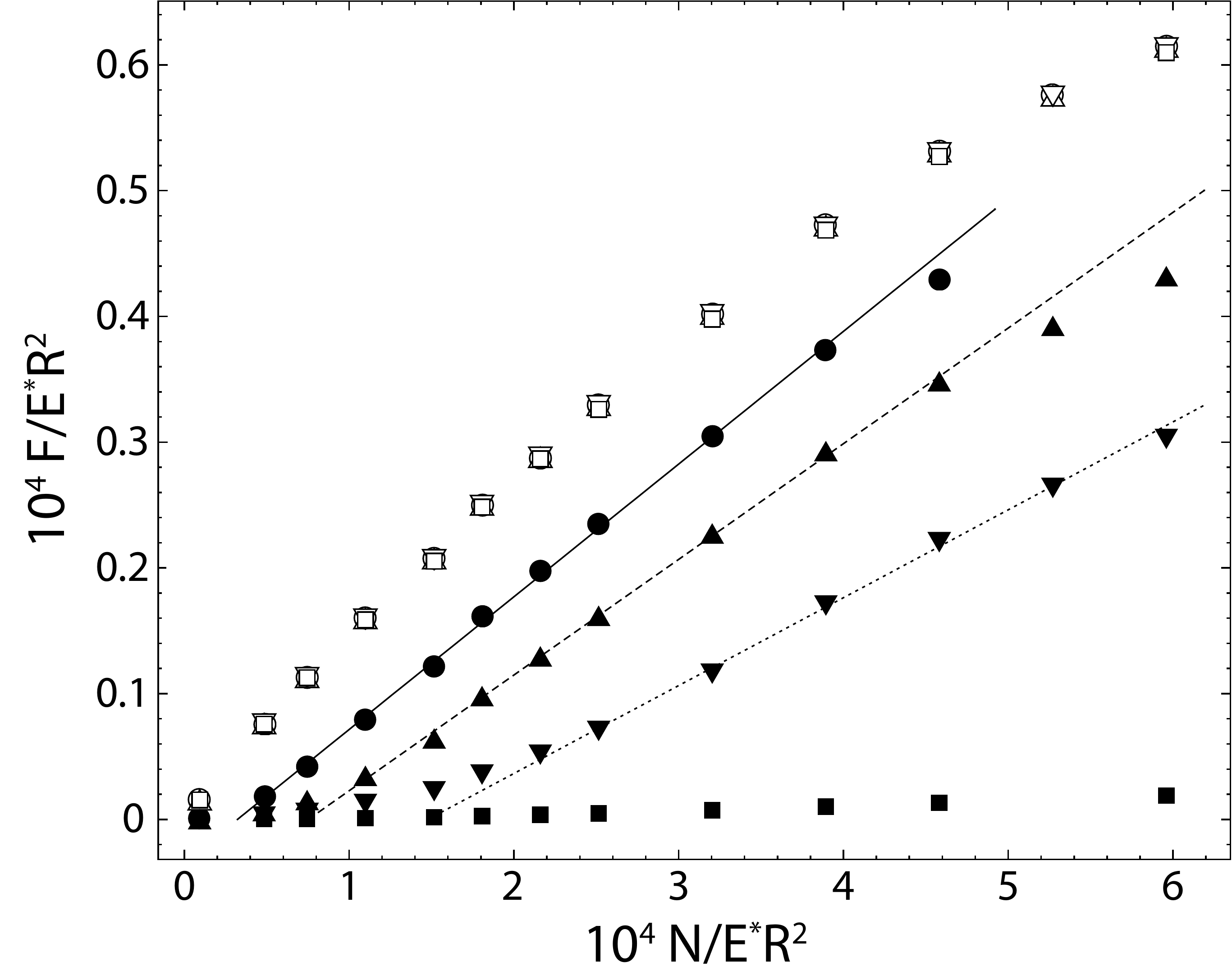}
\caption{Friction on a bare amorphous tip as a function of dimensionless
load at $v=0.01 \sigma/\tau$ and $k_B T/\epsilon=10^{-4}$.
The static friction (open symbols) and kinetic friction (filled symbols)
are shown
for constant tip velocity (squares) and springs with stiffness
$40 \epsilon/\sigma^2$ (downward triangles),
$20 \epsilon/\sigma^2$ (upward triangles),
and  $10\epsilon/\sigma^2$ (circles).
}
\label{BareAmorph}
\end{figure}

Figure~\ref{BareAmorph} contrasts the static and kinetic
friction on a bare amorphous tip over the range of
dimensionless loads considered by Mo et al..
The static friction is strongly sublinear as noted by Luan and Robbins.
The kinetic friction is consistent with the results of Mo et al..
As expected from the Prandtl-Tomlinson model, the friction is 
nearly zero at low loads where the driving system is stiff compared
to the substrate potential.
At larger loads the friction rises roughly linearly with load.
The crossover between these behaviors moves to larger loads as the
stiffness of the pulling spring increases.
Note that the sensitivity to stiffness is lower when an adsorbed
film is present because the potential energy is not a simple
function of tip position as assumed in the Prandtl-Tomlinson model.

The results in Fig.~\ref{BareAmorph} explain the apparent discrepancy
between previous results.
Luan and Robbins did not report $F_k$ because of the strong dependence
on stiffness.
Mo et al. did not report $F_s$ or vary the system compliance by
adding a spring.
However, the tip, substrate and hydrogen bonds terminating the tip
all act as compliant elements.
The importance of grafted hydrogen and hydrocarbons in locking surfaces
together and dissipating energy has been shown in a number of studies by
Harrison and coworkers~\cite{harrison93,harrison99}.

\begin{figure}[htb]
\centering
\includegraphics[width=3in]{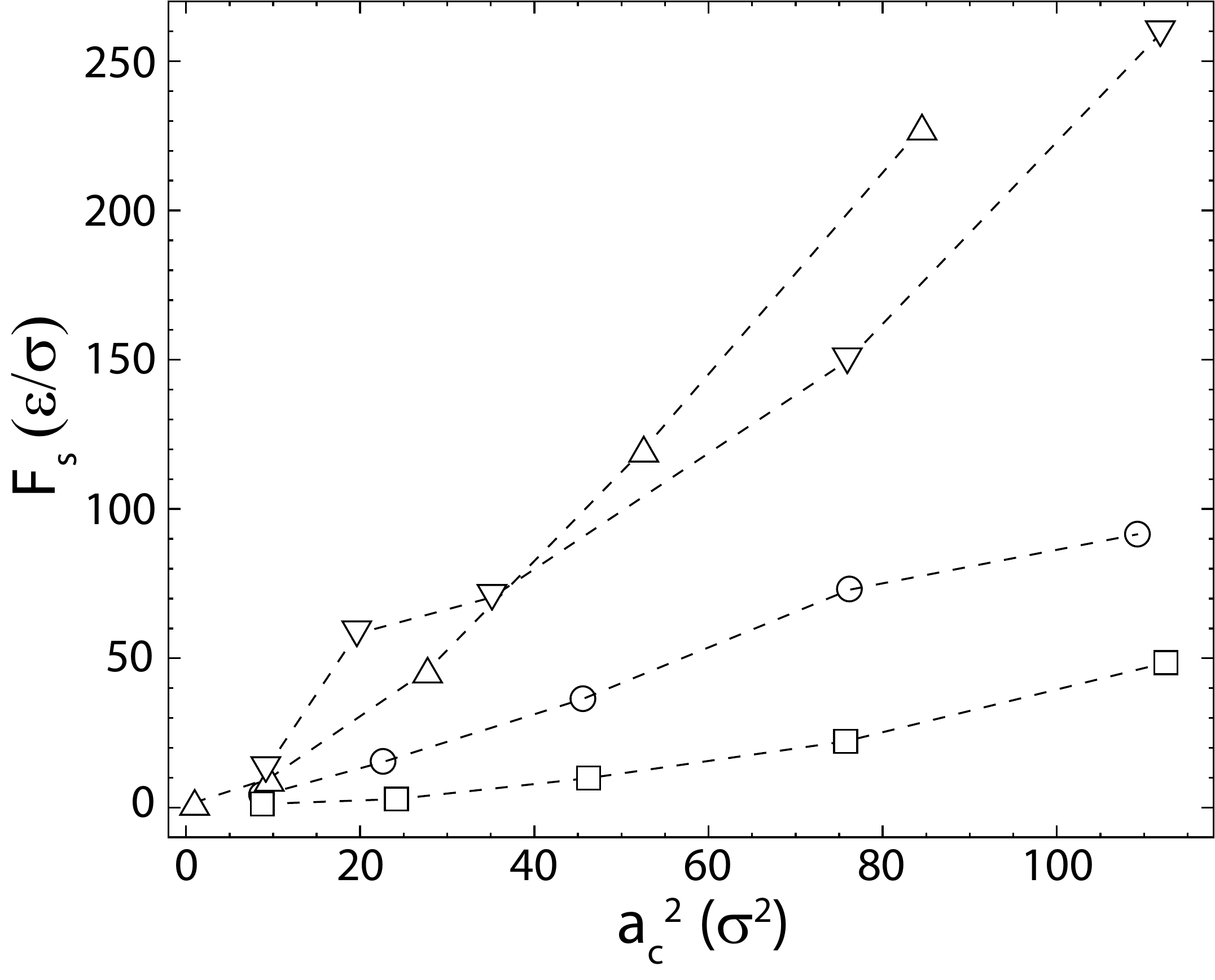}
\caption{Static friction from Fig. \ref{fric}(a) vs. contact radius squared
$a_c^2$ for the indicated tip geometries:
amorphous ($\bigcirc$), incommensurate ($\square$), 
out-of-registry commensurate ($\bigtriangleup$), 
in-registry commensurate ($\Diamond$) and stepped ($\bigtriangledown$).
}
\label{FrictionArea}
\end{figure}

Both previous studies of bare tips
found that the friction rose roughly linearly with area as measured
either by $a_o$~\cite{luan05,luan06b} or $a_c$~\cite{szlufarska09}.
However, the friction did not necessarily got to zero at zero area,
and in many cases the outer region contributed little to the friction.
Fig.~\ref{FrictionArea} shows a plot of friction vs. $a_c^2$ for
tips sliding on a monolayer.
The behavior at large forces could be fit to a line that reaches
zero friction at a nonzero area.
However, the friction is clearly not zero below this point and results
at lower area show pronounced curvature.
The curvature is even more pronounced when $a_o$ is used instead of $a_c$
as would be expected from the fact that $a_o$ scales 
as load to the $1/3$ power with an offset.
We conclude that when a film is present the friction force is more strongly
connected to normal load than area of contact.

\section{Summary and Conclusions}

A single fluid monolayer was shown to have a significant impact on the
mechanical properties of single-asperity contacts.
Although the molecules diffused rapidly on an isolated substrate, they entered
a glassy state when confined under the tip.
As a result, the film was only expelled at the highest loads
and for an in-registry commensurate tip where the barriers to
motion were smallest.
Previous simulations of hydrocarbons between flat surfaces have also
found films remain stable to very high pressures \cite{landman96,persson00}.
The pressure gradient produced by spherical tips should facilitate
the expulsion of the film, but it remained stable over the times
available to our simulations ($\sim 1 \mu$s).
The loads used are comparable to the largest in AFM experiments
and produced surface strains of more than $10\%$, which would
lead to plastic indentation for
most potentials other than the ideal springs considered here.

Sliding at high load was more likely to damage the film,
but the disruption rapidly healed through fluid diffusion
once the tip had passed.
Sliding only produced 
direct contact for the stepped tip and only at the highest load.
The in-registry commensurate tip expelled the film under static
loading, but lifted on top of the film during sliding.
All the above results suggest that relatively weakly adsorbed
molecules may remain within contacts under typical loading
conditions.
Their ability to separate opposing surfaces and recover from damage
are desirable features for boundary lubricants.

The adsorbed film spreads the contact pressure over a substantially
larger area (Figs.~\ref{WetDry} and \ref{ContactRadius})
than for bare tips or Hertz theory.
Deviations in the variation of pressure with radius between different
tip geometries were smaller than for bare tips because the film
smeared features out over a molecular diameter.
This smearing also explains the large increase in the outer radius
of the pressure distribution relative to Hertz.
Values of $a_o$ are shifted by a constant of order the radius
where the tip surface has risen above its lowest point by a
molecular diameter, $a_{\sigma} \equiv \sqrt{2\sigma R}$.
This is a significant fraction of the radius for typical AFM
tips, but would be a small effect in the larger contacts
of a Surface Force Apparatus.

The pressure in the outer regions is small and $a_o$ may overestimate
the effective size of contacts.
Contact radii derived from moments of the pressure distribution are
closer to the values for bare tips and to Hertz theory.
At large loads, $a_s$ is shifted from Hertz theory by a constant offset
of a few molecular diameters.
As $N$ decreases, $a_s$ saturates.
In this regime the local contact pressure is comparable to the ideal gas
pressure.
The tip is supported by thermal fluctuations in the upper edge
of the monolayer.
The width of this layer is of order $\sigma$ and the pressure is
spread over a radius of order $a_{\sigma}$.

The contact area was also determined from the number of tip
atoms in contact at any instant multiplied by the area per atom.
The corresponding radius $a_c$
is just slightly below the Hertz prediction for most tips and loads.
The source of the large discrepancy between $a_c$ and $a_s$ at low loads
is that each tip atom is only in contact for a small fraction of the time.
This is consistent with the tip being supported by thermal fluctuations
in the upper edge of the monolayer.

A surprisingly large temporal variation in the identity of the atoms
that make contact
was also observed at high loads.
The number $N_c$ of tip atoms that contacted the adsorbed
film over a time interval $t$ grew roughly as $t^{1/4}$ 
while the number contacting a bare substrate grew as $\ln t$.
For bare surfaces the time dependence appears to reflect
thermal vibrations in the substrate.
For the adsorbed film, slow displacements of molecules to new
glassy states and fluid fluctuations in low pressure regions
at the edge of the contact also increase $N_c$.
The time dependence of $N_c$ makes it a more ambiguous measure
of contact area than measures based on the mean pressure.
Even increasing $t$ to the period of the fastest phonon vibrations
changed the contact area by $50\%$.
Purely repulsive interactions were used here, but an attractive
tail in the potential also complicates the definition of contacting
atoms.

The adsorbed film also increases the normal tip displacement $\delta$,
which is directly accessible in some experiments~\cite{kiely98}. 
The in-registry commensurate tip exhibits a large jump in displacement
when the film is expelled.
For other tips the displacement rises at the rate
predicted by continuum theory at large loads.
At smaller loads $\delta$ rises more rapidly.
This is qualitatively consistent with continuum theory for
a thin elastic coating where the effective modulus crosses over
from that of the coating to that of the more rigid substrate
as the contact radius becomes much larger than the film
thickness~\cite{johnson01,sridhar04,adams06}. 
The deformation of the substrate is very close to Hertz theory,
but is slightly reduced because the pressure is spread over a
greater area.

The normal displacement is closer to Hertz theory than the contact
radius because
it reflects an average response of the entire system, while
the contact pressure is a local property.
Fits at large loads give accurate values of the substrate modulus,
indicating that analogous experimental measurements provide
accurate material parameters.

The static and kinetic friction were studied for tips moved at
constant velocity or pulled by a spring.
The static friction is insensitive to the 
stiffness of the system, but the kinetic
friction decreases as the stiffness increases (Fig. \ref{fric}).
Similar behavior is observed in the
Prandtl-Tomlinson model~\cite{prandtl28,tomlinson29,muser03acp}
and previous simulations of tips moving over a fixed potential
\cite{socoliuc04,glosli93,gyalog95}.
The rate of decrease is greatest for commensurate tips
where atoms can move coherently between metastable states
and the tip feels a nearly fixed potential.
Amorphous tips show the least decrease because they can not
jump to an equivalent minimum.
The resulting rearrangements in the film lead to greater dissipation
and a larger $F_k$.

For all tips, the static friction is roughly proportional to $N$ at
low loads.
Nonlinear behavior sets in at high loads when there is a transition, such
as contact of a new terrace on a stepped tip or disruption of the film.
The kinetic friction follows the same trends and is slightly smaller
than $F_s$ when the tip is pulled through a weak spring.
For a stiff system, the friction is extremely small for a range
of loads and then rises linearly.
The extent of the low friction region is largest for commensurate
systems because of the coherent motion discussed above.
The value of $F_k$ in this region goes to zero with decreasing 
velocity.
A similar low friction region has been observed in AFM experiments at
low loads and is consistent with 
the Prandtl-Tomlinson model~\cite{prandtl28,tomlinson29,muser03acp}.
The apparent discrepancy between previous results for friction
on bare amorphous tips~\cite{luan05,luan06b,szlufarska09}
could be explained by the fact that one reported $F_s$ while
the latter reported $F_k$.

The relation between friction and contact area was also examined for
each of the definitions described above.
The best correlation was obtained for the static friction and $a_c^2$
(Fig.~\ref{FrictionArea}),
but the dependence is less linear than for load, and appears to
go to zero at finite area.
Note that $a_c^2$ scales roughly as $N^{2/3}$ (Fig.~\ref{ContactRadius}).
If friction is linear in load, plots of friction against area should
have the upwards curvature observed in Fig.~\ref{FrictionArea}.

The linear relation between friction and load is consistent with
previous simulations of adsorbed layers between flat surfaces.
For a wide range of atomic structure and surface coverage
the shear stress $\tau_{shear}$ is a linear function of
the local pressure: $\tau_{shear} = \tau_0 + \alpha p$.
If this linear relation persists between rough surfaces, then the
total friction
should scale as $F = \tau_0 A_{real} + \alpha N$ for any distribution
of pressure.
For the case of nonadhesive interactions considered here $\tau_0$ is nearly
zero and friction should rise linearly with load as observed.
Moreover a direct comparison of the values of $\alpha$ for simulations
of flat surfaces with the ratios $F_s/N$ observed here shows
the same trends with structure and similar numerical values.
For example, commensurate surfaces yield $\alpha \approx 0.3$,
incommensurate surfaces have $\alpha \approx 0.05$, and amorphous
surfaces have $\alpha \approx 0.1$~\cite{he99,he01b,muser01prl}
Simulations also show that $\alpha$ is insensitive to the exact
coverage~\cite{he01b}, explaining why damage to the film does
not produce dramatic changes in $F/N$.

Studies of thin films between surfaces with a small random roughness found a
similar linear rise in $\tau_{shear}$~\cite{landman04}.
The authors argued that the real area of contact was irrelevant in
these simulations because
there was no direct contact of opposing surfaces, but any definition
based on contact pressure would give the real area of load proportional
to the apparent area in their simulations.
It would be interesting to extend simulations to much rougher 
surfaces where the contact pressure is only significant over 
a small fraction of the apparent area and where this $A_{real}$
rises with load.

Another important extension will be to consider a monolayer which partially
wets the tip.
In this case a capillary meniscus will form around the tip.
In addition to producing an adhesive force on the tip, the
capillary may introduce new dissipation mechanisms.
It may also serve as a reservoir that feeds molecules to the
contact and prevents film rupture.
Another parameter that could be varied is the thickness of the film
on the substrate.

\section*{Acknowledgments}
This material is based upon work supported by the National Science Foundation
under Grant No.~DMR-0454947 and the Air Force Office of Scientific
Research.


\end{document}